\documentclass{aa} \usepackage{graphicx} \usepackage{txfonts} \usepackage{lipsum} 
\usepackage{amsmath}
\usepackage{lscape} \usepackage{placeins} \usepackage{longtable} \usepackage{natbib}
\usepackage{hyperref}
\usepackage{xcolor} 
\bibpunct{(}{)}{;}{a}{}{,}

\makeatletter
\let\linenumbers\relax
\let\@linenumbers\relax

\makeatother

\begin{document} 
\title{Orbital Parameters for  64  Wide Binary Systems: no evidence for gravitational anomaly at low acceleration }
 \subtitle{} 
 \author{L. Pasquini\inst{1} \and R. Saglia\inst{2} \and F. Patat\inst{3} \and L. Berni\inst{1,4} \and D. Bossini\inst{5,6} \and L. Magrini\inst{1}  \and H.-G. Ludwig\inst{7} \and M. T. Murphy\inst{8} \and J.R. de Medeiros\inst{9} \and B.L. Canto Martins\inst{9} \and J. Chanam\'e \inst{10} \fnmsep\thanks{Based on ESO-VLT observations, programme 116.28NN.001} } 
 \institute{INAF, Osservatorio Astrofisico di Arcetri, Largo E. Fermi 5, 50100 Firenze Italy \email{luca.pasquini@inaf.it} 
 \and Max Planck Institute for Extraterrestrial Physics, Giessenbachstr. 85748 Garching, Germany 
 \and ESO, Karl Schwarzschild Strasse 2, 85478 Garching bei München, Germany 
  \and Dipartimento di Fisica e Astronomia, Università degli Studi di Firenze, Via Sansone 1, 50019, Sesto Fiorentino, Italy
 \and Dipartimento di Fisica e Astronomia, Università degli studi di Padova, Vicolo dell'Osservatorio 3, Padova, Italy 
 \and INAF, Osservatorio Astronomico di Padova, Vicolo dell'Osservatorio 5, Padova, Italy 
 \and Landessternwarte Zentrum f\"ur Astronomie der Universit\"at Heidelberg, K\"onigstuhl 12, 69117 Heidelberg, Germany
 \and Centre for Astrophysics and Supercomputing, Swinburne University of Technology, Hawthorn, Victoria 3122, Australia 
 \and Departamento de Fisica, Universidade Federal do Rio Grande do Norte, 59078-970 Natal, RN, Brazil 
 \and Instituto de Astrofísica, Pontificia Universidad Católica de Chile, Av. Vicuña Mackenna 4860, 782-0436 Macul, Santiago, Chile}
\date{Received September 30, 20XX} 
\abstract {Wide binaries have been recently studied by  several groups to test gravity in the low-acceleration regime, where deviations from Newtonian dynamics may appear. } 
{We aim to understand whether wide binaries, when analyzed with state of the art tools (that include accurate radial velocities and selection for multiple systems, common proper motion pairs and  chance pairs) can be represented by bound Newtonian orbits, and, if this is the case, whether the distribution of the orbital parameters retrieved agrees with expectations.} 
{In order to explore the range of possible orbits consistent with the observations and the orbital parameters' distribution, we added a Monte Carlo analysis to our previous study. The potential orbits  are classified according to the $\chi^2_{tot}$ probability of representing the observed quantities: star coordinates, parallaxes, velocity components and  masses. We apply this analysis to 44 systems  with accurate radial velocity measurements previously  published, to which we add 25 wide binaries  from literature, consisting of twin low-mass (late M main sequence) stars, also with multiple, fairly precise radial velocity measurements.  } 
{We find orbital solutions for 64 systems, and derive  the distribution of their orbital parameters.
   We find that for most wide binaries the parameter distributions are too complex to be properly represented
  by one (best fit)  value and an associated standard deviation.  We also find that the single  best model (the one with the lowest $\chi^2_{tot}$)  is not always representative of the
  overall distribution. Consequently, we conduct our analysis by aggregating the normalized probability distributions rather than relying on individual best-fit values.
  For one wide binary system composed of M-type stars, no Newtonian bound solution can be found because the velocity difference between the two components is excessively large;
  however, this pair is a member of the young Argus association. This binary should not be considered a suitable target, as the stars are likely a common proper motion pair
  that will eventually drift apart. The remaining four systems that cannot be fit, were already identified as likely disrupted binaries. The uncertainty in the separation between the components, given by the
    parallax error, is by far the dominant error source, and the main limitation of the present analysis,  affecting especially the determination of  the orbits' inclination angle  and  phase.}  
{ Newtonian dynamics describe well our sample, there is no need  for new physics and there is no evidence for excess velocity difference  at low accelerations. 
Orbital parameters distributions for the whole sample agree with theoretical expectations.  
When dividing the sample into two subsamples, according to the mutual acceleration, the eccentricity and periastron distributions agree with theoretical expectations;
  inclination and phase are distorted by the parallax-limited separation, more strongly in the high-acceleration (37 systems) than in the low-acceleration ($A_N < 10^{-9} m s^{-2}$, 27 systems)
  subsample, which is opposite to what a gravitational signal would produce. This is independent of the A$_N$ threshold chosen.
 In particular the ratio between observed velocity difference and circular velocity does not show any excess for the
  low acceleration sample. When parametrized as a boost $\gamma$ parameter to the velocity, $0.93 < \gamma <1.03 (1 \sigma)$, fully consistent  with Newton also  for the low acceleration sample. 
Finally, our sample is well represented by a thermal eccentricity distribution (f(e)=2e); much larger  but less constrained surveys find evidence for super-thermal eccentricity for separations  larger than  1 KAU. }

\maketitle 
 
\section{Introduction}

Detached binaries are the simplest example of a cluster of stars. As such, they have been used in the  past for a number of studies, for instance, to test  stellar evolution models \citep{Barrientos+2021, Espinoza+2021}
or to study the effect of planetary systems on  photospheric abundances \citep{Spina+2021}.  Wide binaries (WBs),  in addition, 
offer a unique opportunity to test gravity in the low-acceleration regime. WBs  in fact experience gravitational accelerations comparable to those in the outer regions of galaxies, however, dark matter does not significantly influence their dynamics, as its density is known to be negligible in the solar neighborhood \citep{Soding+2025}.  
 A further advantage of WBs is that their dynamics is well understood. They can be treated as isolated systems because the differential acceleration exerted by the Galaxy on the two stars is negligible.
Similar low accelerations can be found, for instance, in the outer regions of globular clusters. However, these systems exhibit complex dynamical behaviour, including tidal tails and internal rotation, which makes them poorly suited as probes for testing Newtonian gravity.
  
Excited by the  potential of testing gravity with WBs,  several  works have used the exceptional {\it Gaia} astrometry to construct large samples of WBs with precise projected separations and transverse velocities (hereafter referred to as 2D analyses)  \citep{Pittordis+2019, Banik+2024, Chae+2023, Chae+2024, Cookson+2026}. However, these studies have reached conflicting conclusions on whether Newtonian gravity best reproduces the observations.

Another line of research  focuses on the study of 3D solutions, by adding   precise and accurate radial velocities to the {\it Gaia} data. 
High-quality radial velocities are essential for identifying and removing multiple systems from the samples (e.g., cases where a third body orbits one of the WB stars), and, when combined with {\it Gaia} DR3 parallaxes and proper motions, they allow the orbits of binaries to be constrained \citep{Saglia+2025}.
These authors analysed  HARPS \citep{Mayor+2003} archival spectra for 32 WBs, finding that 31 of them could be represented  with a viable bound Newtonian solution. Radial velocities have also been used by \citet{Chae2025a}, who selected {\em Gaia}'s best-quality radial velocities and a Bayesian approach to compare Newtonian and MOdified Newtonian Dynamics (MOND) solutions, finding that the latter are preferred for separations beyond 2 kAU. Similar conclusions were reached by \citet{Chae2025b}, who applied a Bayesian analysis to the HARPS sample published by \citet{Saglia+2025}.  A similar analysis was repeated for a new sample of 36 low acceleration (acceleration  $< 10^{-9} ms^{-2}$)  WBs \citep{Chae+2026} finding that for all  systems, irrespective of their acceleration,  Newtonian gravity is not the preferred solution. Finally, \citet{Pasquini+2026} analysed a sample of 12 wide binaries with large separations (greater than 13,000 AU) and found that for three systems no bound Newtonian orbits were possible. However, these binaries have extremely large separations and are therefore expected to be disrupted by interactions with the Galaxy or by encounters with other stars \citep{Hamilton+2024}.
In their study, \citet{Pasquini+2026}  pointed out that low mass systems could be ideal targets because these binaries experience low accelerations for much smaller star separations, 
making therefore the probability of being disrupted by interactions  with Galactic tides or third bodies negligible. 

In this study  we aim at improving our previous analysis in two fundamental aspects.  Previously we focused on finding optimal orbital solutions -- i.e., those that best reproduce all observables -- allowing the distance to vary within the {\it Gaia} parallax uncertainties. However, we did not explore the full uncertainty space or the potential correlations between the orbital parameters. We have implemented a Monte Carlo analysis exploring the full parameter space of possible solutions, and we quantify the results using a $\chi^2$ statistic that accounts for all velocity vectors and the uncertainties in velocities, distances, and masses.
With such an analysis it is possible not only to determine the most probable solutions but also to study the distribution of the orbital parameters.  
We apply this analysis to the 44 WBs with accurate radial velocities that we previously published in \citet{Saglia+2025} and \citet{Pasquini+2026}. 

The second  improvement  is to add  a sample of 26  low mass M-type star binaries published recently  \citep{Couperus+2025a, Couperus+2025b}.These binaries, in addition to being low-mass systems, consist of nearly identical stars and have multiple, rather precise radial velocity measurements. Since the two stars of each WB  have  the same magnitude and colours,  we can use the published radial velocities without the need of differential corrections for gravitational redshifts and convective shifts, because the stars are so close to being real twins  \citep{Couperus+2025a, Couperus+2025b}.  

Even if limited, a sample of 69 WBs provides a reasonable basis to start investigating not only whether WBs have a Newtonian bound orbital solution but also whether the distribution of the retrieved orbital parameters agrees with our expectations.

The paper is structured as follows: In Sections~\ref{sec_2} and 3 we discuss the new M star sample and the revision of the \citet{Pasquini+2026} sample respectively, while the Monte Carlo analysis is explained in Section~\ref{sec_3}, and with further details in the Appendix. Section~\ref{sec_4} illustrates our results, while in Section~\ref{sec_5} we mostly compare them with other recent studies. The last Section (Sect.~\ref{sec_6}) summarizes our conclusions. 

\section{ The M stars binary sample }
\label{sec_2}

In this section we describe the sample of M-dwarf wide binaries used in our analysis.
\citet{Couperus+2025b, Couperus+2025a}  studied in detail a set of 40 nearby M dwarf binaries. Although the authors  primarily focussed on stellar activity,
they also provided a set of radial velocity measurements for a subsample of 27 WBs. These measurements are 
suitable for orbital analysis. We excluded one binary, KX Com A +BC because BC has variable radial velocity, and a second system,
 composed of  RTW0824-3054A+B, because the B component is likely a binary itself \citep{Couperus+2025b}.

In their recent work, \citet{Pasquini+2026} argued that low-mass stars could be optimal targets for gravity tests because, for a given acceleration,
their separations are much smaller than those of the most massive stars.
Since the main parameter governing perturbations from stellar encounters is the orbital semi-major axis, these systems are also less sensitive to such effects.

A key characteristic of this sample is that the components of each binary are `twins': the two stars not only have similar parallaxes and proper motions, but also
differ by less than 0.1\,mag in {\it Gaia} BP and RP colours, as well as in 2MASS J, H, and Ks magnitudes.
This is an essential property for our  analysis, because  we aim at the highest radial velocity precision that can only be obtained 
when the instrumental  velocity of the stars is corrected for gravitational redshift and  convective blue shift. 
Since these two quantities can be as high as many hundreds of m\,s$^{-1}$, they are not negligible  compared to a difference in radial velocity of a few hundred
m\,s$^{-1}$ typical of a binary in the low acceleration regime.  
Since these M binaries are composed of true twins,  we can use the  instrumental  radial velocity difference without any further correction. 
\citet{Couperus+2025b} provide, in addition to the radial velocities, stellar masses. We adopted the masses, radial velocities and their uncertainties from
\citet{Couperus+2025a, Couperus+2025b}, and the coordinates, parallaxes, proper motions, and their uncertainties from {\it Gaia} DR3 to derive the orbital solutions.  

For one binary (RTW~0933-4353~A+B), the published radial velocity uncertainty (V(A) = 19.70$\pm$ 2.71 and V(B) = 21.78$\pm$ 4.41 km s$^{-1}$)
was exceedingly large, so we observed the two stars with the ESPRESSO spectrograph at the VLT \citep{Pepe+2021}. In our measurements the relative radial velocity of the two stars
(V(A)-V(B) = -2.24 km s$^{-1}$) is consistent with   \citet{Couperus+2025b};  however, our absolute values differ from the Couperus' ones by approximately
3~km s$^{-1}$. Given the  uncertainty in the ESPRESSO radial velocity is of $\sim 80$ m s$^{-1}$, this difference between ESPRESSO and \citet{Couperus+2025b} is relevant,
as will be discussed later. Consequently, we use for this sample the data by \citet{Couperus+2025a} and \citet{Couperus+2025b},
but we adopt our ESPRESSO radial velocities for stars RTW~0933-4353~A and RTW~0933-4353~B (13.28$\pm$0.08 and 15.52$\pm$0.07~km\,s$^{-1}$ respectively).

\section{The 'Chaname' sample \citep{Pasquini+2026} }
  
 In revising our sample, we discovered that the stellar parameters of five   binaries from the \citet{Pasquini+2026} sample were not correct.
We therefore recomputed the stellar parameters for the entire sample using the methodology described in that work, but incorporating the improvements
developed in the companion paper (Berni et al. 2026).  An errata corrige for the \citet{Pasquini+2026} paper is also in preparation. In a nutshell, the previous analysis was converging
in a secondary minimum close to the solar parameters, and this affected all the stars with high effective temperatures.
In particular, the companion paper introduces a differential analysis of the binary components, designed to achieve substantially
higher precision in the derived chemical abundances. As a consequence, the spectral and stellar evolution analyses were repeated for all systems.

The revised analysis leads to significant (that is, larger than the quoted original uncertainties) changes in the stellar masses and radii for at least one component of five  binaries.
For all the remaining systems, the differences are negligible and do not affect the orbital solutions presented in \citet{Pasquini+2026}.
We report the updated stellar parameters for the five binaries
in Table B1, keeping the original \citet{Pasquini+2026} numbering.

\section{Methods}
\label{sec_3}
In this section, we describe the methods used to model the geometry of the systems and to derive the orbital parameters of the wide binaries.

The geometry of the systems, as well as the equations and the orbital parameters are described  in \citet{Saglia+2025} and \citet{Chae+2026}, while \citet{Leao+2019} discuss the details of the radial velocity calculations.

The method used in \citet{Saglia+2025} and \citet{Pasquini+2026} aimed to reproduce exactly the observed velocities and the observed $\Gamma_{2D}$ and $\Gamma_{3D}$ angles,  
varying the distances between the stars within the {\it Gaia} uncertainties and, if needed, the stellar mass.
Only the $\chi^2$ on the distance was used to quantify the agreement between the observed quantities and the model.
This method is efficient and fast for finding solutions, but it does not fully explore the solution space,
nor does it provide a robust estimate of the uncertainties of the retrieved orbital parameters.

To test whether Newtonian physics can reproduce the observations of WBs, it is not sufficient to show that WBs can be fit with bound orbits;
we must also assess whether the distribution of orbital parameters is consistent with theoretical expectations.
A complete analysis should therefore include not only a fit to bound Newtonian orbits but also a verification
that the orbital parameters behave as expected, along with a physically motivated explanation in cases where they do not.

In order to investigate the orbital parameter space we determine the orbital elements of each pair by a Monte Carlo procedure:
the observational uncertainties are propagated by drawing a large number of realisations of the uncertain observables, solving the orbit exactly for each one.
No optimisation is involved at any stage, because the orbit is determined analytically once the observables are fixed.

For each pair we determine the relative orbit from the observed state vector
rather than by fitting. Given the distances $D_1,D_2$, the total mass
$M=m_1+m_2$, and the relative position and velocity
$(\mathbf{R},\mathbf{V})$ constructed from the astrometry and the radial
velocities, the six Keplerian elements $(a,e,i,\Omega,\phi_0,M_0)$ are fixed
uniquely and follow in closed form from the standard vector relations for the
specific energy, the specific angular momentum $\mathbf{h}=\mathbf{R}\times
\mathbf{V}$, the node vector $\mathbf{n}=\hat{\mathbf{z}}\times\mathbf{h}$ and
the eccentricity vector (see appendix  A for the detailed description). 

All vectors are expressed in the local North--East--line-of-sight frame
$(\hat{\mathbf{e}}_\delta,\hat{\mathbf{e}}_\alpha,\hat{\mathbf{e}}_{\rm los})$,
whose axes are the ICRS unit vectors in the directions of increasing
declination, increasing right ascension, and the line of sight away from the
observer, evaluated at the position of the primary.  With this convention
$i<90^\circ$ denotes motion towards increasing position angle and $\Omega$ is
the position angle of the receding node.  We retain retrograde solutions
($i>90^\circ$): the only exact degeneracy of an astrometric orbit,
$(\Omega,\phi_0)\rightarrow(\Omega+\pi,\phi_0+\pi)$, is broken by the radial
velocities.
Because $(a,e,i,\Omega,\phi_0,M_0)$ are determined exactly, the measurement
uncertainties are propagated by sampling rather than by optimisation.  We draw
$N$ realisations of the six observables that carry uncertainty --- the two
parallaxes, the total mass, and the three components of the relative velocity
--- \emph{uniformly} over a box of half-width $n_\sigma=3$ in units of each
standardised residual $z_k$, and solve the orbit for every realisation.  The
statistic
\begin{equation}
  \chi^{2}_{\rm tot}=\lvert\mathbf{z}\rvert^{2}
  =\chi^{2}_{D_1D_2}+\chi^{2}_{M}+\chi^{2}_{V}
\end{equation}
is then the squared distance of that realisation from the observed quantities (See appendix A, equation  \ref{eq:chi2tot}).  Sampling
uniformly, the weight $\exp(-\chi^{2}_{\rm tot}/2)$ must be considered in the following analysis.
 It should be noticed that $ \chi^{2}_{\rm tot}$ differs substantially from zero only when the input values of D$_1$,D$_2$ and M  do not yield a
  bound elliptical solution; it measures therefore the smallest displacement in masses and distances required to obtain one Newtonian solution.
  It is a measure of tension with the input values, not a fit residual.

The implementation is vectorised and verified at run time: solutions are
propagated back through a forward Keplerian model, reproducing the sampled
state vectors with a maximum relative residual of $\sim\!10^{-12}$, and the
elements have been validated against direct numerical integration of the
two-body problem, agreeing to $<10^{-6}$~deg in all angles. 
 The orbital solutions are saved in tables that can later be analysed.

\section{Results } 
\label{sec_4}
\begin{figure}
\centering
 \includegraphics[angle=270, width=8.0cm]{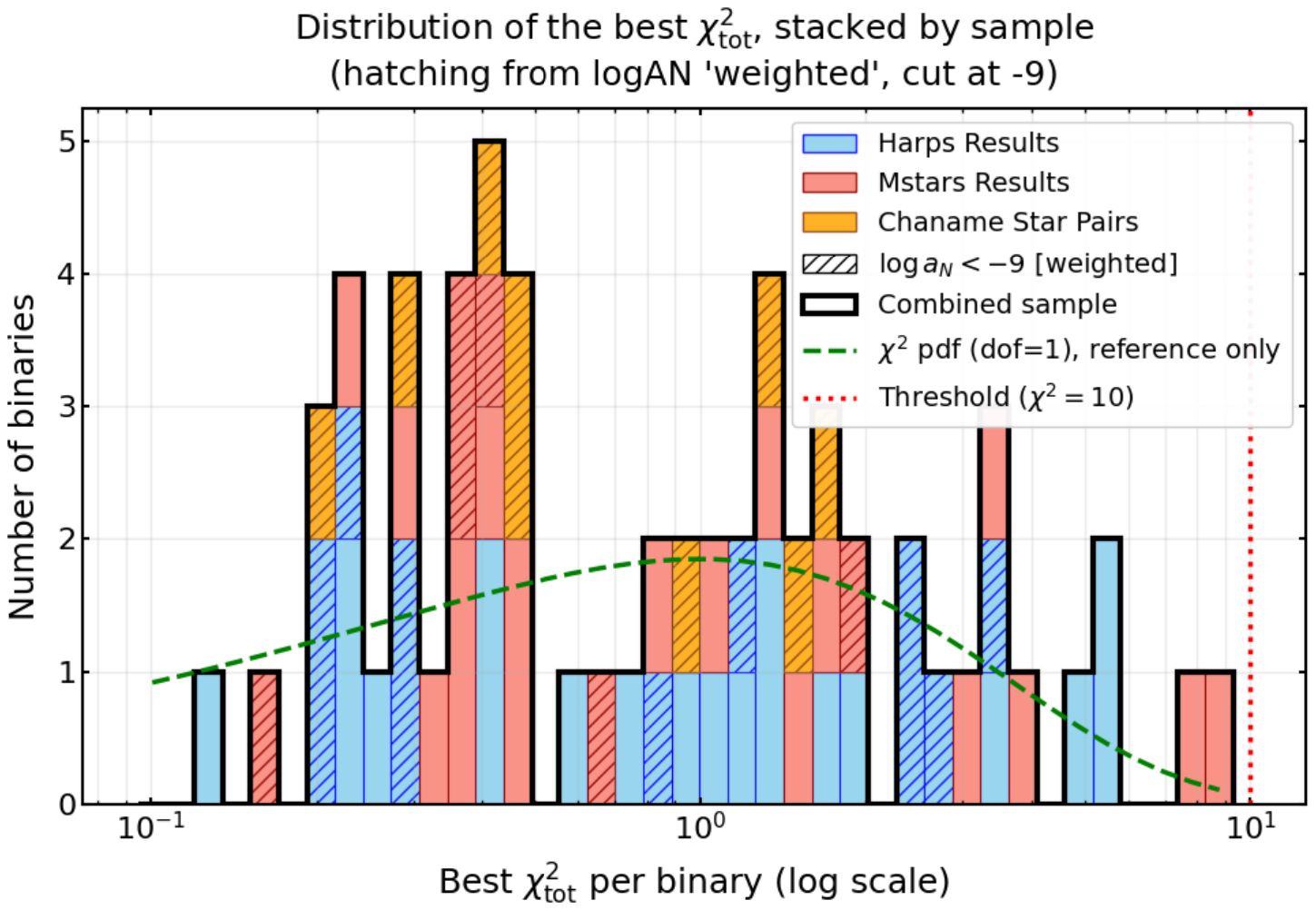}
   \caption{Minimum $\chi^2_{tot}$ distribution for the 64 WBs with bound Newtonian orbital parameters. Each subsample is marked with a different colour. WBs belonging to the 'low acceleration' subsample are marked with dash lines  }
         \label{Fig1}
  \end{figure}

The MC analysis provides solutions (and associated $\chi^2_{tot}$)  for 64  of the 69 WBs studied: 31 from the \citet{Saglia+2025} sample,
9 from the \citet{Pasquini+2026} sample and 24 from the M sample. 
The WBs for which no acceptable solution could be found are the same as those previously identified in the literature: WB~24 in \citet{Saglia+2025}, WBs~7, 11, and 12 in \citet{Pasquini+2026},
plus the pair of M-stars  RTW~0933-4353~A+B, which is discussed in more detail below. 
Figure \ref{Fig1} shows the  distribution of  the minimum $\chi^2_{tot}$ determined for the 64 WBs with bound Newtonian orbital solutions.
Binaries belonging to the different samples are marked with different colours, and the theoretical $\chi^2$ distribution is superimposed.  The results closely follow the $\chi^2$  distribution.
 
\begin{figure*}
\centering
 \includegraphics[angle=270, width=16.0cm]{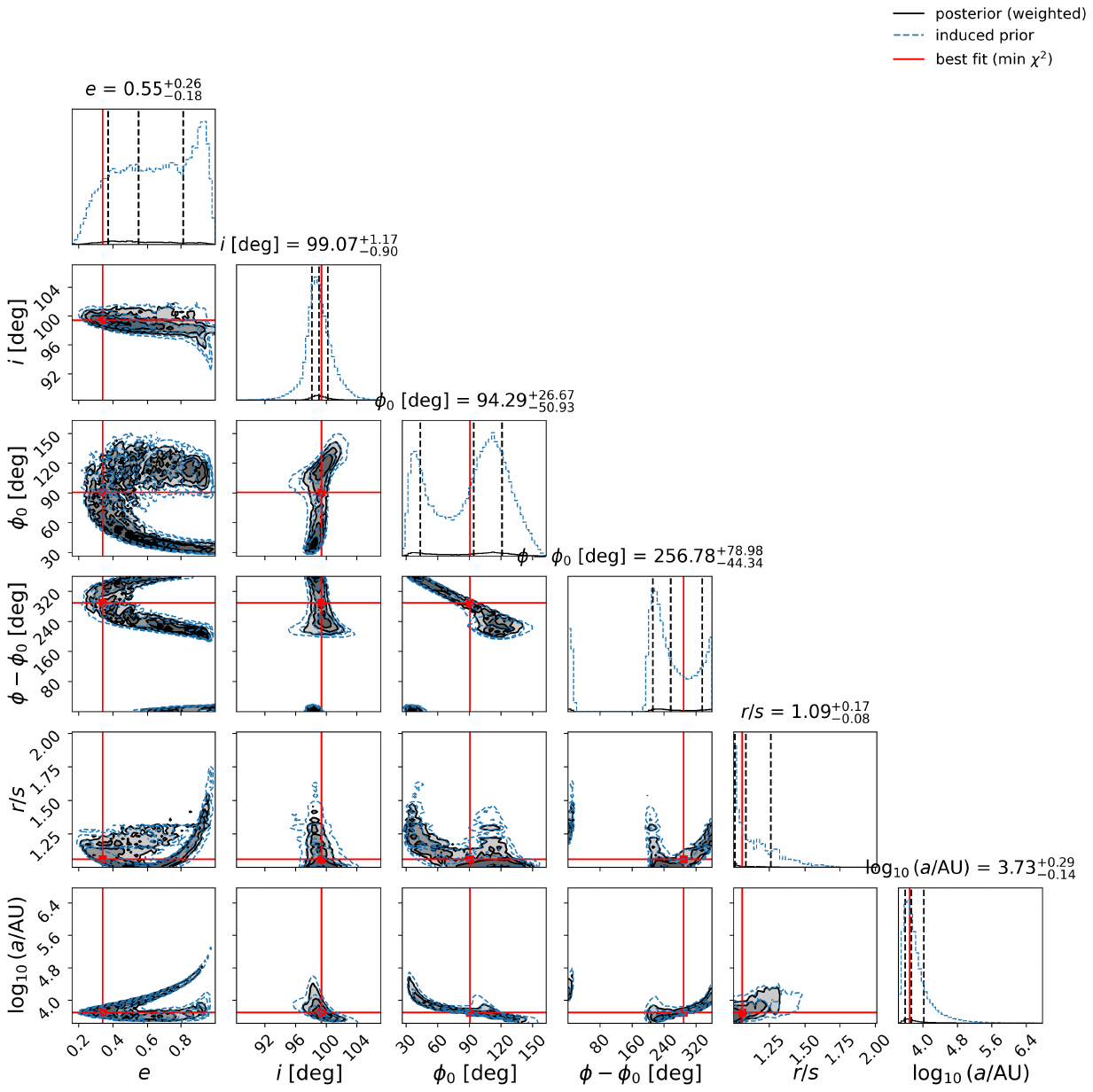}
 \caption{The distribution of orbital parameters for the  wide binary  number 5 of the HARPS sample (HD13357A+B).
   The vertical red line indicates the values corresponding to the lowest (best) $\chi^2_{tot}$ solution found. This Figure shows how the solutions' parameter space
   might be complex   and cannot be represented by one value with associated uncertainty. }
         \label{Fig2}
  \end{figure*}

\begin{figure*}
\centering
 \includegraphics[angle=270, width=16.0cm]{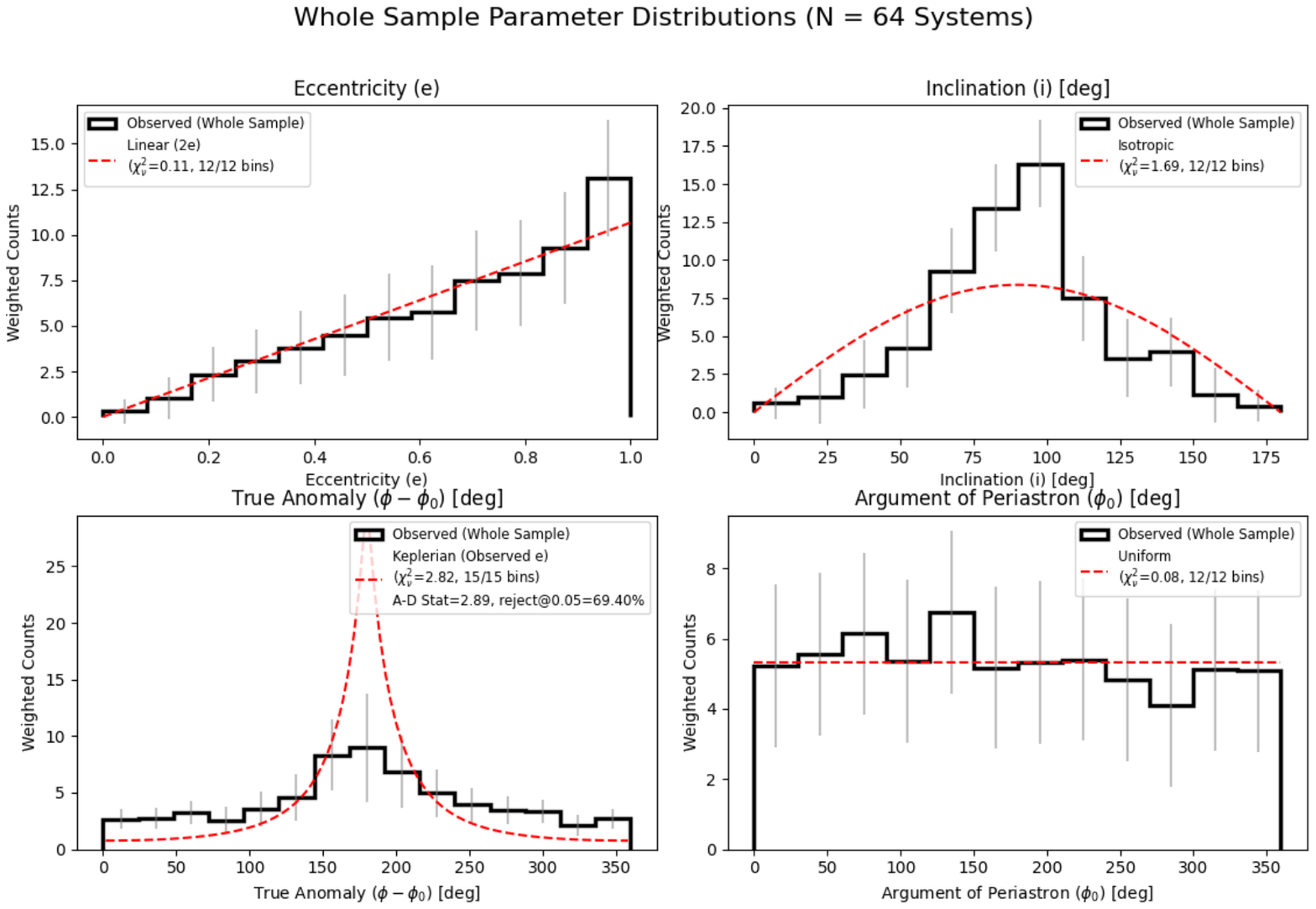}
 \caption{Distribution of the orbital parameters eccentricity, true anomaly, argument of periastron  and inclination for the whole sample. The expected distributions
   are superimposed, and the $\chi^2$ quantifying the agreement is given.   }
         \label{Fig3}
  \end{figure*}

 In Figure~\ref{Fig2} we show one example of the parameter space.  We use the `corner' function in Python
 \citep{foremanmackey2016corner} to visualize the data, using as
 weight for each solution the likelihood $L = e^{-\chi^2_{tot}/2}$.
 For some of the systems, the space of the orbital parameters can be well delimited,
   but for several systems the solution space is complex and cannot simply be
 represented by a central value and an associated dispersion, as shown in the example (HD13357A+B).   In the same
 figure, we plot in red the values of our `best solution', defined as
 the solution with the minimum $\chi^2_{tot}$. It is worth noting that
 the best solution may not correspond to the peak of the parameter
 distribution. 
 
 For these two reasons, it is not appropriate to characterise the orbital parameters with a single value and an associated $\sigma$.
 Instead, for each WB parameter we consider the full distribution of the parameter space obtained from the MC analysis, weighted by the likelihood $L$ (see Figure \ref{Fig2}).
 For each WB orbital parameter, the distribution is normalised by dividing by the sum of the likelihood weights (up to a maximum of $\chi^2_{tot}$ minimum + 12.59),
 so that the integral of each individual distribution equals unity. The normalised distributions are then summed.
 
 The results of this process are shown in Figure~\ref{Fig3} for the
 entire sample of 64 WBs, for the orbital parameters: eccentricity,
 true anomaly ($\phi-\phi_0$), periastron angle $\phi_0$ and
 inclination angle of the orbit ($i$).  The expected scaling laws are
 superimposed: thermal distribution for eccentricity ($F(e)=2e$), true
 anomaly $F(\phi-\phi_0) = (1-e^2)^{1.5} /(2\pi\,(1+e\times\cos(\phi-\phi_0))^2$)), following eq. 13 of \citet{Chae+2026},
 periastron angle (constant) and inclination angle (proportional to
 $\sin(i)$).  Note that the expected distribution for the true anomaly
 depends on eccentricity, and we computed the true anomaly expected distribution using the observed eccentricity distributions.

 The agreement between the expected and observed distributions is
 quantified via a $\chi^2$ computation, assuming that the uncertainty
 in each bin is given by the corresponding Poisson noise of the expected counts.
 
 Clearly, the sample as a whole is well represented by Newtonian
 orbits and a thermal eccentricity distribution. There is no evidence
 of any deviating parameters.

 \subsection{The low and high acceleration cases}

 While the agreement on the whole sample is reassuring about the robustness of the method,
   it does not address the core of the  question on the validity of Newton's law, because it is clear that for high accelerations Newtonian dynamics works in wide binaries \citep{Saglia+2025}.

   Considering,  in addition,  that several theories (MOND, Scaling Invariant Gravity, \citep{Maeder+2024}) expect departure
   from Newtonian physics only at low accelerations, and that this dichotomy is also found in
some recent works \citep{Chae2025b, Chae+2026}, it is important to
investigate whether low acceleration binaries behave differently than
their high acceleration counterparts.  Following \citet{Chae+2026}, we
divided the sample into two subsamples, selecting our WB sample  
according to the mutual Newtonian acceleration A$_N$. Setting
the limit to A$_N =10^{-9}$ m s$^{-2}$  where the acceleration refers to the weighted median of A$_N$ across the solutions.
  We choose this way of selecting because this is the most stable criterion, but
  we tested our conclusions also adopting other selection criteria for $A_N$.
  We have 27 WBs in the low
  acceleration subsample and 37 WBs in the high acceleration subsample.
  In Figures 4-7 we show the orbital parameters, together with their expected curves. In order to quantify the agreement between observations and expectations we use,
  as for the whole sample, a
  $\chi^2$ metric, but we also computed the Anderson-Darling test for the true anomaly ($\phi - \phi_0$) , as suggested by \citet{Chae+2026}.
  We could not implement this test exactly  as suggested by this author,  because we do not have
  one single value of true anomaly per system, but a distribution. We  performed therefore a Monte Carlo resampling, by using 500 iterations.
  Each iteration draws one single orbit
  solution value per system using the likelihood weights and compares it to the synthetic theoretical law. At the end of each iteration,
  \texttt{scipy.stats.anderson\_ksamp} performs a two-sample Anderson-Darling test comparing the array of observed values against the theoretical values.
  The validity of the agreement is quantified in two ways:
a) by computing the mean Anderson-Darling parameter A$^2$, b) by computing the number of solutions whose occurrence probability is less than 0.05.

 \begin{figure*}
\centering
 \includegraphics[angle=270, width=16.0cm]{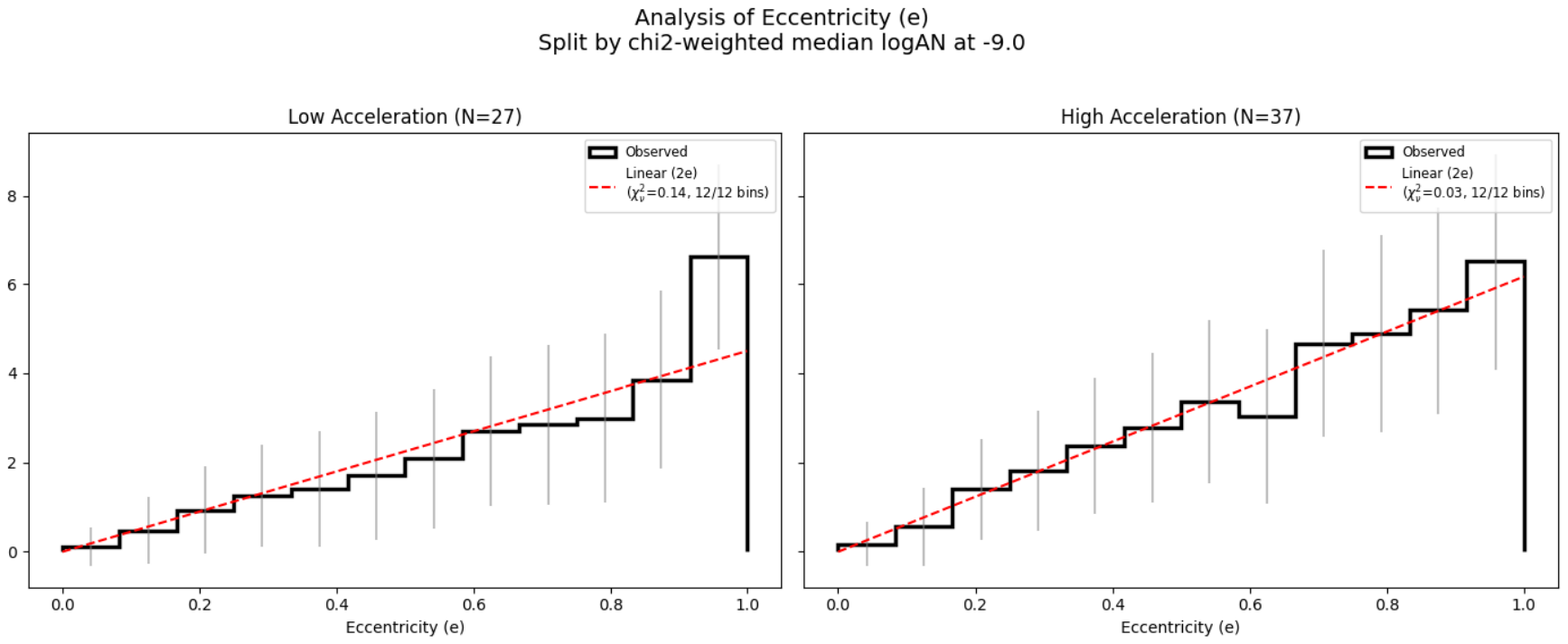}
   \caption{Distribution of the  eccentricity for the two subsamples. The expected distributions are superimposed, and the $\chi^2$ quantifying the agreement is given.   }
         \label{Fig4}
  \end{figure*}

 \begin{figure*}
\centering
 \includegraphics[angle=270, width=16.0cm]{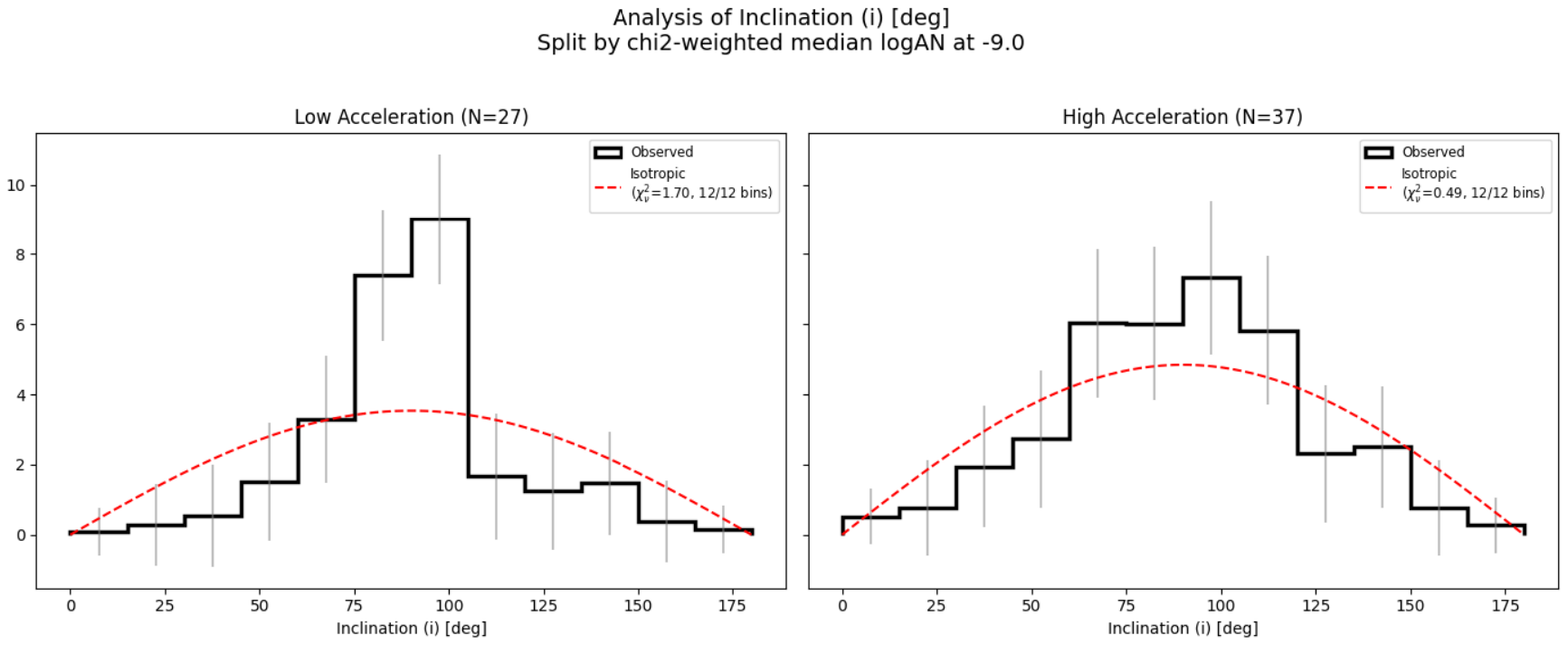}
   \caption{Distribution of the  orbit inclination for the two subsamples. The expected distributions are superimposed, and the $\chi^2$ quantifying the agreement is given.   }
         \label{Fig5}
  \end{figure*}

\begin{figure*}
\centering
 \includegraphics[angle=270, width=16.0cm]{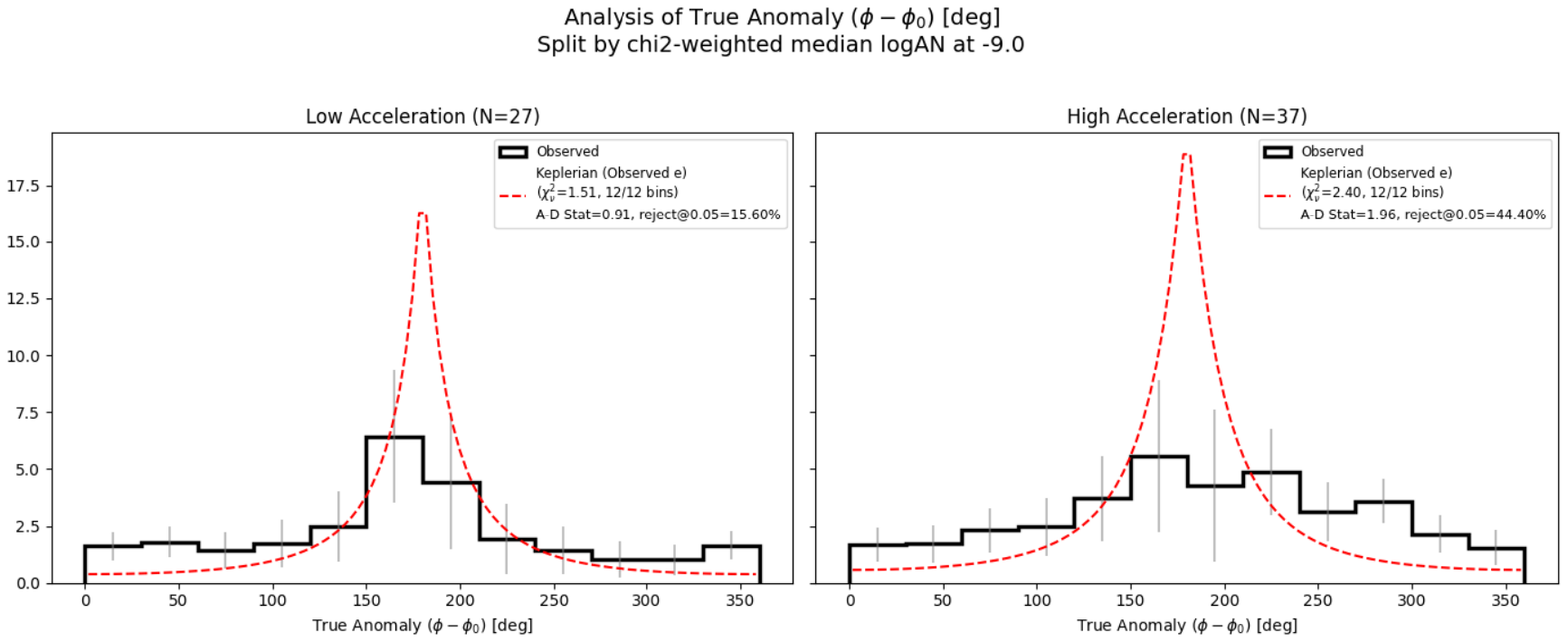}
   \caption{Distribution of the  true anomaly ($\phi - \phi_0$)  for the two subsamples. The expected distributions are superimposed, and the $\chi^2$ quantifying the agreement is given.   }
         \label{Fig6}
  \end{figure*}
  
  \begin{figure*}
\centering
 \includegraphics[angle=270, width=16.0cm]{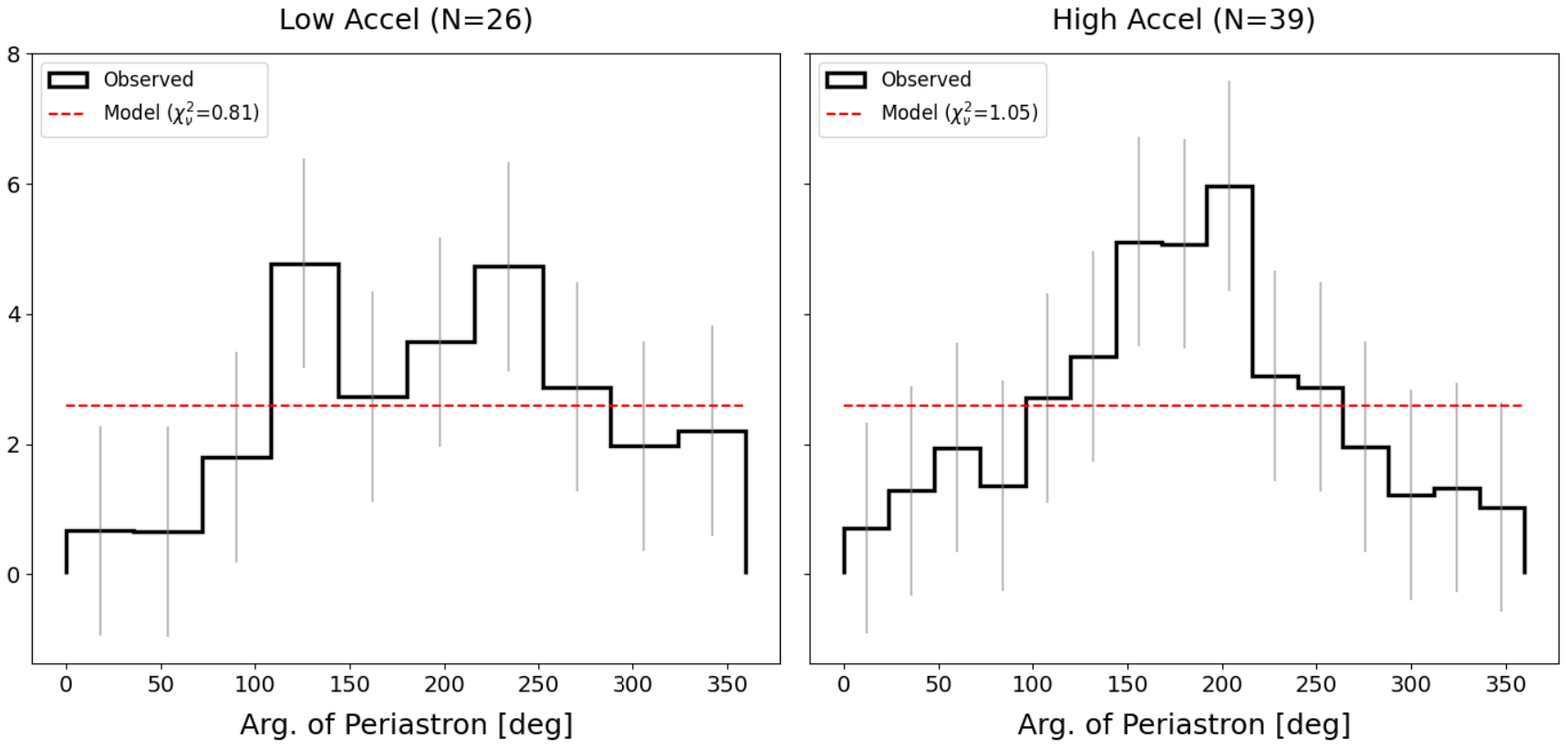}
   \caption{Distribution of the  argument of periastron ($\phi_0$)  for the two subsamples. The expected distributions are superimposed, and the $\chi^2$ quantifying the agreement is given.   }
         \label{Fig7}
  \end{figure*}

All orbital parameter distributions  are consistent with the expectations, with  small  differences  between the two sub-samples.  \citet{Saglia+2025} and \citet{Chae2025b} record an excessive number of WBs close to pericenter (with true anomaly close to 0 or 360).
This excess is potentially quite interesting, because it could show an excess of velocity (difference) in these systems.
 We also observe a flattening in the distribution, but we attribute it to the  uncertainty in separation rather than to a physical excess. 
We have tested different binning, finding that the total $\chi^2$ remains constant, independent of the binning, so  the reduced $\chi^2$ increases with coarser bins.
  This indicates that the discrepancy is localized rather than distributed across the range. We attribute it to the large uncertainty on the 3D separation:
  the observed distribution is the intrinsic one convolved with a broad error kernel, which smears out the peak.
  What remains constant for all binnings is that the low acceleration subsample agrees with the expectations,
  while the high acceleration does not, which is the opposite of what would be expected from a gravitational signal. The imperfect agreement with the expectations in the high acceleration regime
  is almost certainly due to the uncertainty of the 3D separations. While a definitive answer requires forward simulations,
  which are beyond the scope of this paper, we notice that for the high-acceleration subsample the ratio between the uncertainty on the line-of-sight separation, $\sigma(D_1-D_2)$,  and the projected separation $s$ has a median value of 31, while this is 10 for the low-acceleration ones. 
The tighter, high-acceleration pairs are therefore those whose orbital phase is least constrained.  

We also note that this result does not depend on the chosen cutoff for the acceleration: by lowering  the A$_N$  threshold  to A$_N <10^{-9.5}$ m s$^{-2}$,
a value closer to the MOND typical acceleration value,  all values remain consistent with expectations. The low acceleration sample becomes smaller
(e.g. only 19 WBs remain in this subsample); the same is true lowering further the threshold, but less systems remain in the low acceleration bin.

The orbital inclination also deserves a comment. The peak around 90 degrees does not represent the true inclination distribution of the orbits.
  The argument is subtle, but it originates in the large uncertainty on the 3D separation r between the binary components, which is dominated by the parallax error. 
  Since  $cos(i)_{min} = s/r$ \citep{Saglia+2025}, large  values of r  push $i_{min}$ towards 90 degrees, and confine the inclination angle `i' around that value.
  The typical error on parallax would allow r/s up to $\sim$50, but the imposition of bound orbits implies $ r < 2GM/V^2$ , that is equal to 3
  for a typical value of $\tilde{\nu}$ = 0.8.
  The accepted solutions therefore populate r/s between 1 and 3, and this is what concentrates i near 90 degrees.
  We verified this by simulating wide binaries with isotropically distributed
  inclinations at 100 pc with typical parallax errors of 0.02 mas: the retrieved inclinations peak by a factor $\sim$2 too high near 90 degrees,
  and the size of the peak scales with the median r/s. 

As additional consistency check, we examine the expected and observed distribution of 3D-to-projected separations ($r/s$),
since, as discussed in \citet{Pasquini+2026}, high $r/s$ values
 are unlikely and may indicate non-bound or disrupted systems.  In Figure~\ref{Fig5} we plot the $r/s$ values for the full sample and the two subsamples,
finding that they are  rather well represented by the expected distribution 
($r/s = (\cos^2 \phi + \cos^2 i \times \sin^2 \phi)^{-1/2}$).
However, there is an important difference: the median value for the observed r/s is 1.9 and 2.1 respectively for the high
  and low acceleration subsamples, instead of 1.13 as expected by the theoretical distribution.
  This is exactly the boost effect in r mentioned above, responsible for the clustering of the inclination angle around 90 degrees. 

  \begin{figure}
\centering
 \includegraphics[angle=270, width=8.0cm]{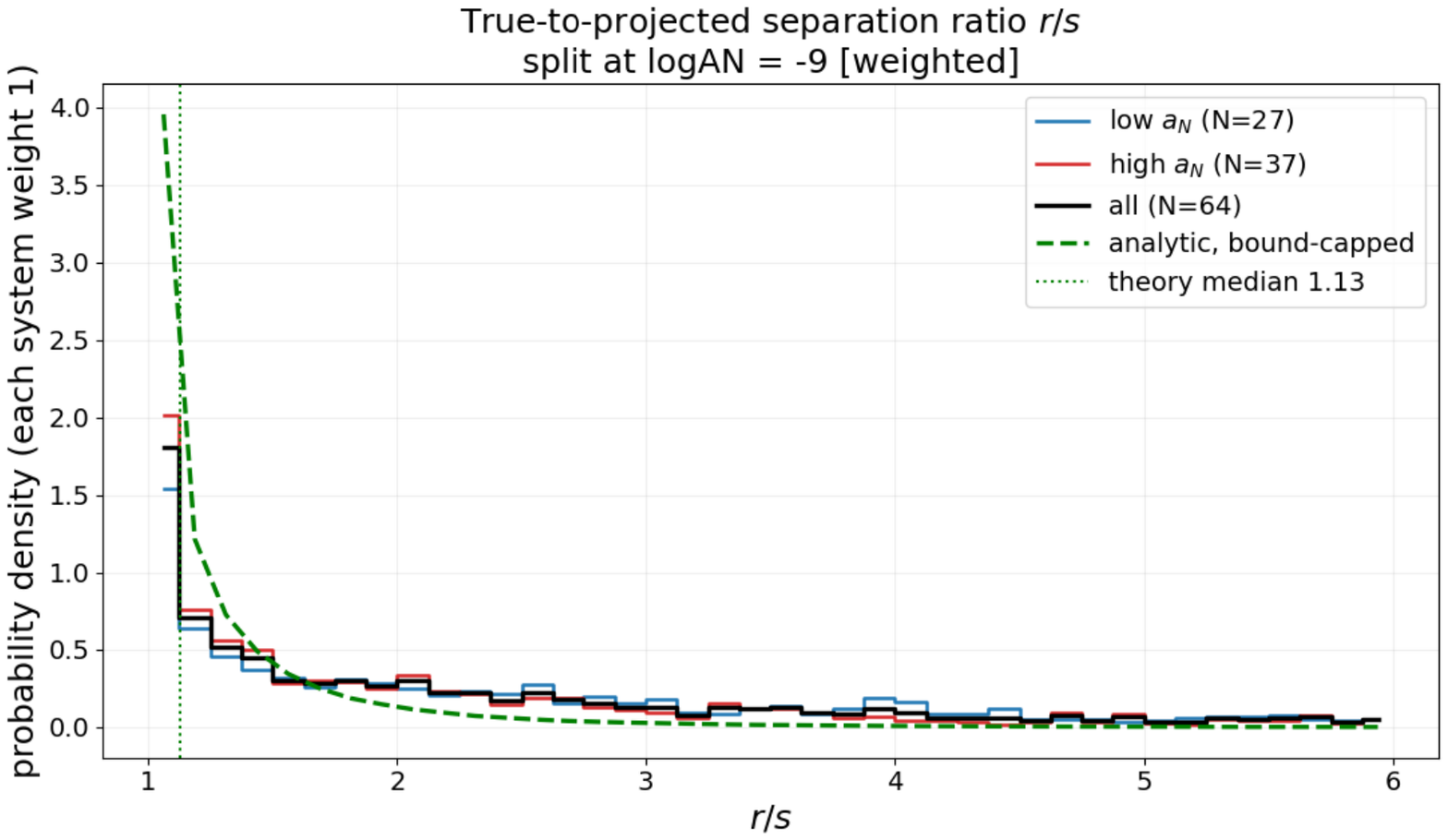}
  \caption{Distribution of the ratio 3D separation and projected separations (r/s) for the whole sample and for the two subsamples with 'high' and 'low' acceleration (limit was set at A$_N = 10^{-9}$ m s$^{-2}$).  The expected distribution is also shown.}
  \label{Fig5}
 \end{figure}

  Since the dominant uncertainty is  the 3D separation between the binary components, it is especially relevant to
    use a diagnostic that does not use this quantity.
  We investigate whether the $\tilde{\nu} = V_{tot}/V_{cir}(s) $ distribution shows any evidence for excess at low acceleration.  This parameter
    has been proposed as diagnostic and simulated for several Newton and MOND cases by \citet{Banik+2018}.  We use the 3D version of the diagnostic, since our
    radial velocity differences are measured  accurately. Most studies, based on 2D analysis,  use the analogue 2D quantity, with only the projected  velocity.
    
    This velocity ratio is very powerful because  all its components are extremely solid observationally:
    the uncertainty associated to  V$_{cir}(s) = \sqrt{GM/s}$ is only a few percent, dominated by the uncertainty on mass. The uncertainty associated to  V$_{tot}$ is typically few tens of ms$^{-1}$
   \citep{Saglia+2025, Pasquini+2026}. 
    
\begin{figure*}
\centering
 \includegraphics[angle=270, width=16.0cm]{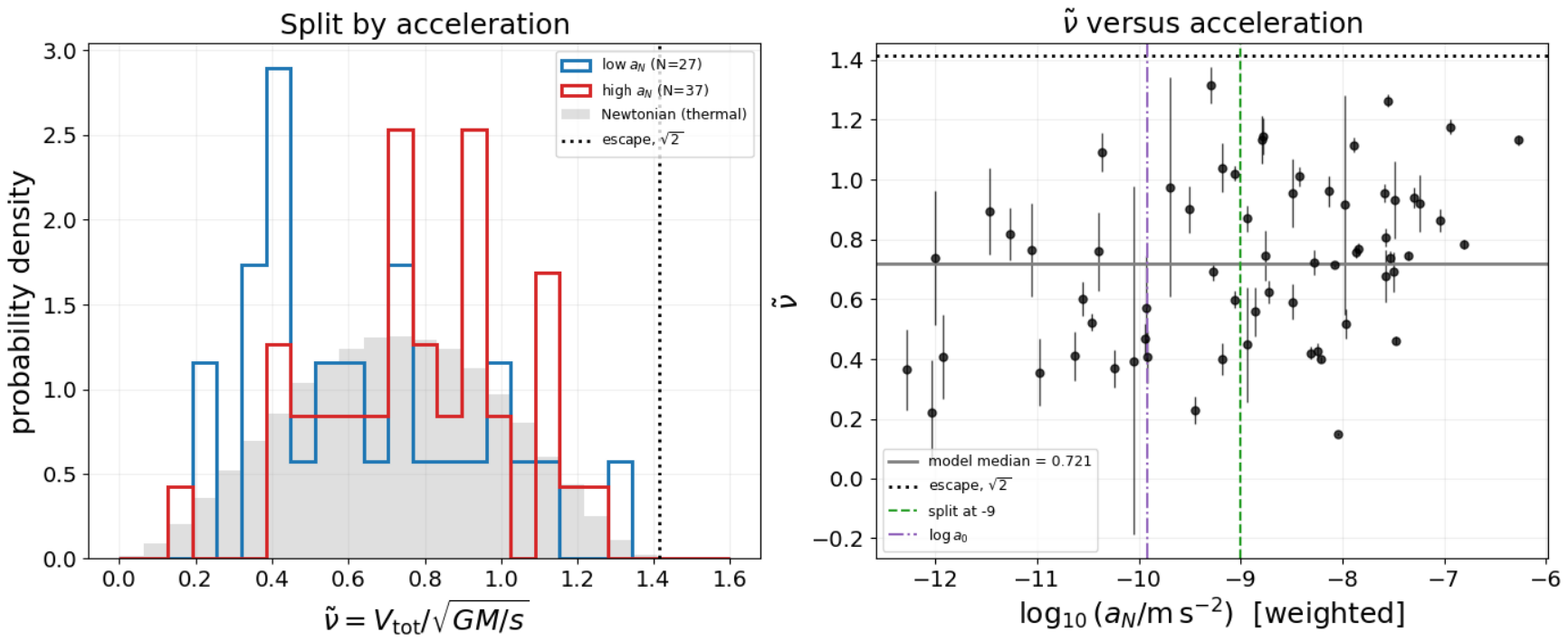}
 \caption{Left: $\tilde{\nu}$ distribution; the expected Newtonian distribution is superimposed; low and high acceleration subsamples binaries are marked with different colours.
   No evidence for enhanced $\tilde{\nu}$ at low acceleration. Right: $\tilde{\nu}$ vs. acceleration. The threshold value of  Log$A_N$ is marked, as well as the typical value of MOND acceleration. }
  \label{Fig5b}
 \end{figure*}

   In  Figure~\ref{Fig5b}  the $\tilde{\nu}$  distribution is shown, with the expected distribution for Newtonian
   gravity and thermal eccentricity. The median values of $\tilde{\nu} $ for the low and high acceleration samples are of 0.60  and 0.77, respectively.
   The plot on the right shows the same quantity  as a function of the Newtonian acceleration Log($A_N$). In this Figure, in addition to the threshold value, a second acceleration close to a classical MOND value, is also shown.  
 
 \subsection{Testing the boost factor}
  
 Even if the sample is small, the $\tilde{\nu}$ test allows us also to probe gravity parametrization, by 
 modifying  gravity increasing the orbital velocities by a factor $\gamma$,
$\tilde \nu\rightarrow\gamma\tilde \nu$, so that $\gamma^{2}=G_{\rm eff}/G$ in terms
of an effective gravitational constant. This comparison is two-sided: values of $\tilde{\nu}$ that are too large
favour $\gamma>1$ and values that are too small favour $\gamma<1$.

For each pair $k$ we form the dimensionless velocity ratio
\begin{equation}
  \tilde \nu_k \;\equiv\; \frac{|\mathbf{V}_{{\rm tot},k}|}
                            {\sqrt{G M_{{\rm tot},k}/s_k}},
  \label{eq:vtilde-def}
\end{equation}
The uncertainty $\sigma_k$ on $\tilde \nu_k$ is obtained by
propagating the parallax, proper-motion, radial-velocity and mass errors
through Equation~(\ref{eq:vtilde-def}) by Monte Carlo, and is the standard
deviation of the resulting distribution.

Under Newtonian gravity $\tilde \nu$ is a purely geometric quantity.
Writing $r$ for the instantaneous separation and $a$ for the semi-major
axis, vis-viva gives
\begin{equation}
  \tilde \nu^{2}=s\left(\frac{2}{r}-\frac{1}{a}\right)
              =p\left(2-\frac{r}{a}\right),
\end{equation}
with $p\equiv s/r=[1-\sin^{2}\phi\sin^{2}i]^{1/2}$ the projection factor of
Eq.~(A.15).  Both $GM_{\rm tot}$ and the orbital scale cancel.
$\tilde \nu$ therefore depends only on the eccentricity $e$, the inclination
$i$, the argument of periastron and the orbital phase, and a single
universal sample serves every pair regardless of its mass or
separation.

We define $P_{\rm N}$  the probability distribution of $\tilde \nu$
obtained  when the orbital elements are drawn from their population distributions: $\cos i\sim\mathcal{U}(-1,1)$
(isotropic orientation), $\phi_0\sim\mathcal{U}(0,2\pi)$, mean anomaly
$M\sim\mathcal{U}(0,2\pi)$ (uniform in time, i.e.\ Kepler's second law), and
$e$ thermal.
For each draw Kepler's equation is
solved for the eccentric anomaly, giving $\phi-\phi_0$ and $r/a=1-e\cos E$.

We denote by
\begin{equation}
  \{u_{j}\}_{j=1}^{N_{\rm MC}} \sim P_{\rm N}
\end{equation}
a Monte Carlo sample of size $N_{\rm MC}=2\times10^{4}$ drawn from this
distribution: each $u_{j}$ is the value of $\tilde \nu$ that one randomly
oriented, randomly phased Newtonian orbit would present to the observer.

The single free parameter $\gamma$ rescales the predicted velocity,
$\tilde \nu_{\rm pred}=\gamma u$, so that $\gamma=1$ is Newtonian and
\begin{equation}
  \gamma \;=\; \sqrt{G_{\rm eff}/G_{\rm N}}
\end{equation}
for a pseudo-Newtonian modification.

Each model prediction is convolved with that pair's own measurement
uncertainty, so the likelihood of $\gamma$ is

\begin{equation}
\begin{gathered}
  \ln L(\gamma)=\sum_{k}\ln\left[\frac{1}{N_{\rm MC}}
    \sum_{j=1}^{N_{\rm MC}}
    \mathcal{N}\!\left(\tilde \nu_{k};\gamma u_{j},\sigma_{k}\right)\right],\\[4pt]
  \mathcal{N}(x;\mu,\sigma)=\left(2\pi\sigma^{2}\right)^{-1/2}
    \exp\!\left[-\frac{(x-\mu)^{2}}{2\sigma^{2}}\right].
\label{eq:gamma-like}
\end{gathered}
\end{equation}

The inner sum is the Monte Carlo estimate of the model density at
$\tilde \nu_{k}$, i.e.\ of $\int \mathcal{N}(\tilde \nu_{k};\gamma u,\sigma_{k})
P_{\rm N}(u)\,{\rm d}u$.  The same sample $\{u_{j}\}$ is drawn once and reused
for every pair and every trial value of $\gamma$, so that differences in
$\ln L$ across the grid reflect the model and not sampling noise.

Equation~(\ref{eq:gamma-like}) is evaluated on a uniform grid
$\gamma\in[0.3,4.0]$ of step $0.01$ and maximised, giving $\hat\gamma$.  
Since $\gamma$ is the only free parameter, the confidence interval is read
directly off the likelihood curve: we quote the range of $\gamma$ over which
$\ln L$ falls by no more than $\Delta\ln L=0.5$ below its maximum, which
corresponds to $1\sigma$ for a single parameter ($\Delta\chi^{2}=1$).
 
The outcome is that the boost $\gamma$ factor is 0.98 (0.93-1.03 1$\sigma$ interval)
for the low acceleration subsample and $\gamma$ is 1.07 (0.99-1.10 1$\sigma$ interval)
 for the high acceleration one. Again, by using different threshold values for
 $A_N$ , the situation remains similar with the $\gamma$ boost factor actually slightly decreasing for the low acceleration subsamples.
 The referee asked us to include two systems considered disrupted by
\citet{Pasquini+2026}, HARPS \#24 and Chanam\'e \#12, on the grounds that their
projected separations might not be  large. With these two systems the low-acceleration
subsample grows from 27 to 29 pairs, and a single boost factor fit gives
$\gamma = 1.17^{+0.09}_{-0.08}$. This value is driven by the two added pairs, which have
$\tilde\nu = 2.23$ and $1.63$, i.e. $\tilde\nu > \sqrt{2}$: a scaled Newtonian
distribution can only accommodate them by stretching $\gamma$ for the whole
subsample.

A boost of the gravitational constant, as predicted by MOND, shifts the whole
$\tilde\nu$ distribution, whereas perturbed or unbound pairs produce a few
outliers above an unshifted distribution. To distinguish the two cases we fitted
a mixture model \citep{Hogg+2010}, in which the likelihood of each pair is
\begin{equation}
  p(\tilde\nu_k \mid \gamma, f) = (1-f)\, p_{\rm N}(\tilde\nu_k \mid \gamma)
                                 + f\, p_{\rm out}(\tilde\nu_k),
\end{equation}
where $p_{\rm N}$ is the Newtonian distribution scaled by $\gamma$ and
convolved with the measurement error of the pair (the model used above),
$p_{\rm out}$ is an outlier component, uniform in $\tilde\nu$
between 0 and 4 and convolved with the same error, and $f$ is the fraction of
outliers. The mixture is strongly preferred over the single boost factor
($\Delta 2\ln L = 11.2$ for one additional parameter, $p = 4\times10^{-4}$).
Its Newtonian core gives $\gamma = 0.89^{+0.15}_{-0.05}$, consistent with
Newtonian dynamics, with a fraction  $f = 0.14^{+0.09}_{-0.10}$,
in the outlier component. The result is stable against the choice of the outlier component:
changing its upper limit from 3 to 5 changes $\gamma$
between 0.88 and 0.98. For the high-acceleration subsample no outlier component
is needed ($f = 0$, $\gamma = 1.07^{+0.03}_{-0.08}$), and for the whole sample
$\gamma = 1.02^{+0.06}_{-0.03}$ with $f = 0.03$.
Including these two binaries therefore does not change our conclusions: the low-acceleration pairs are consistent with Newtonian dynamics once a small fraction of outliers is allowed for.
That outliers appear only at low acceleration is expected in Newtonian dynamics, since these are the widest and most weakly bound pairs,
the most easily perturbed by the Galactic tide and stellar encounters. In addition,  the presence of contaminating, undetected triple systems cannot be excluded.

 \citet{Banik+2018} estimate that at
 least 180 systems are needed to
 disentangle  between the different solutions, so we can only say that the present sample does not require modified gravity, though we cannot exclude it.
 We note that our uncertainties are statistical only, for instance they
 do not consider the systematic effects produced by assuming a distribution of eccentricity different from thermal. 

The need for a stronger gravity could also manifest as a systematic 
increase of mass of the systems in the orbital solutions.  
Since we allow the masses to vary within their expected uncertainty
range (typically about 5$\%$), we verified whether our solutions have systematically higher masses than the input values. This
is not the case: the median ratio  between orbital and input masses is 1.001 with a 0.01 standard deviation and no trend is present with acceleration. We therefore do not introduce any systematic
bias in the derived accelerations.

\subsection{The M stars }
\label{sec:mstars}

The M stars binary sample includes 25 binaries. 
For all the binaries
a bound Newtonian solution can be found, with the exception of  RTW 0933-4353 A+B, for which the
difference in velocity is far too large.  The orbital parameters
for the best solutions (minimum $\chi^2_{tot}$) for the 24 systems are
given in Table~\ref{tab:velocities} and \ref{tab:orbital_params} in
the Appendix.

For RTW~0933-4353~A+B, our ESPRESSO observations yield a radial velocity difference of more than 2~km~s$^{-1}$,
consistent with \citet{Couperus+2025b}, and exceeding the maximum velocity compatible with a bound system. \citet{Couperus+2025b}
discussed this binary as a possible member of the young ($\sim$50 Myr) Argus association, although membership was not confirmed due
to marginal tension with the expected velocity distribution. The two stars are the faintest in their sample, and the CHIRON spectra had low signal-to-noise ratio, as reflected in the large uncertainties.
Our ESPRESSO data, with a precision better than 80 m~s$^{-1}$, were analysed using the BANYAN $\Sigma$ tool \citep{Gaigne'+2025}, yielding membership probabilities in the Argus association of 89.7\% and 27.3\% for the two components, assuming a radial-velocity uncertainty of 0.2~km~s$^{-1}$. The radial velocities were obtained via cross-correlation with an M5 mask, although the absolute zero point of this procedure is not well constrained for these stars.
To assess the impact of this offset, we compare with Proxima Cen: using the same M5 mask, ESPRESSO yields a radial velocity of $-21.36$~km\,s$^{-1}$, about 340 m\,s$^{-1}$ redder than the value of $-21.70$~km~s$^{-1}$ obtained from a detailed analysis by \citet{Kervella+2017}. Applying a similar zero-point correction and adopting an uncertainty of 0.4~km~s$^{-1}$ increases the Argus membership probabilities for RTW~0933-4353~A+B to 90\% and 37\%, respectively.
We therefore conclude that this system is likely a member of the young Argus association and is not suitable for our gravity test. This is particularly unfortunate, as it is the lowest-mass system in the sample.
  
Two additional M-dwarf binaries (RTW~2311-5845~A+B and RTW~2011-3824~A+B) exhibit the highest total $\chi^2_{\mathrm{tot}}$ values (7.97 and 9.11, respectively). In both cases, the high $\chi^2_{\mathrm{tot}}$ is driven exclusively by the distance term ($\chi^2_D$), indicating that the {\it Gaia} parallax uncertainties must be stretched to accommodate the orbital solutions. Future {\it Gaia} DR4 data will clarify whether these solutions are viable.
We note that, according to our orbital solutions, these binaries have accelerations of $A_{\rm N} \sim 10^{-8.7}$ and $10^{-7.2}$ m s$^{-2}$, respectively. They therefore do not lie in the acceleration regime where effects from MOND \citep{Chae+2026} or scale-invariant gravity \citep{Maeder+2024} are expected to be relevant.
 
 \section{Discussion}
\label{sec_5} 
Keeping in mind that our sample of low acceleration wide binaries is
limited to 27 objects, we can say that Newtonian dynamics provides a
{\it coherent} view of the observations.  64 out of the 69 WBs
observed can be represented by bound Newtonian orbits and the orbital parameters retrieved follow the expected
distributions.
 
Of the remaining 5 that cannot be fit, one is convincingly a member of the
young Argus association, so not suitable for the test. For the
remaining 4, \citet{Pasquini+2026} discussed that they are likely
disrupted systems, in view of their large separations or large $r/s$
ratios. These systems are expected, according to secular evolution
models based on Newtonian dynamics, to be disrupted by Galactic tides
or stellar encounters \citep{Hamilton+2024}.
 
 Our results are in apparent contrast  with those found by the analysis of
 \citet{Chae+2026}. These authors found, even in a limited sample of
 36 selected low acceleration (acceleration below 10$^{-9}$ m
 s$^{-2}$) wide binaries, evidence that non-Newtonian gravity should
 be preferred to Newtonian. This result is true for all their systems,
 irrespective of their gravitational acceleration.
  
 We note that there are a number of fundamental differences between
 our work and that of \citet{Chae+2026}:
 
 \begin{enumerate} 

\item The two works aim to answer different questions: We pose the
  question `can the observations be fit with bound Newtonian orbits?'
  \citet{Chae+2026} instead parametrise gravity as a free parameter and pose
  the question `what is the best form of gravity to represent the
  observations'?

 \item We use only radial velocities that are corrected for
   gravitational redshift and convective blue-shifts, unless, as for
   the M star sample in this work, the stars are real twins.
   \citet{Chae+2026} use a mixture of radial velocities, some
   corrected, but the majority not corrected for these quantities.  In
   principle, not applying these corrections should not introduce
   large systematic effects (unless the radial velocity difference A$-$B
   has a preferential sign), but the uncertainty in radial velocity
   should be much larger than the measurement errors. 
   When considering that the non-Newtonian gravitational signal is
   small, this is an important difference. Finally, the calculation of gravitational redshift remains non-trivial in non-Newtonian frameworks. 
   Furthermore, as convective blueshift corrections are highly sensitive to the specific methodology and instrumentation used for radial velocity measurements, 
   they cannot be reliably extrapolated to other datasets unless the full details are known and the zero point corrections
   are calibrated \citep{Leao+2019}.
 
 \item An additional difference between the samples concerns the selection criteria. In this work (as in our previous studies), we apply an age cut: young systems (typically younger than 1 Gyr) are excluded, as are pairs belonging to young clusters and associations. This is to exclude common proper motion
   pairs and unbound binaries.
 
 \end{enumerate} 
 
In general, the differences between the two analyses are so substantial that a direct comparison between our work and that of \citet{Chae+2026} is currently not possible, though we cannot support \citet{Chae+2026} conclusions.

One point can be explored in more detail: \citet{Saglia+2025}
and \citet{Pasquini+2026} found 4 WBs that cannot be fit with bound
Newtonian orbits, and \citet{Pasquini+2026} argued that they are
disrupted binaries, because of their large separations or of their
large $r/s$ ratio. In their work \citet{Chae+2026} found three
additional low acceleration systems (systems 3, 33, 42, cf.  their
Table 3). Their system 59 is not a new identification, as it corresponds to pair 7 in \citet{Pasquini+2026}. The ratio between the most probable 3D and
projected separations ($r/s$) for these three new systems ranges from
$\sim$3.9 for pair~3 to $\sim$1 for pair~42, and therefore does not
present any strong anomaly.  Of these 3 systems, one (pair 42),
composed of HD 95238 and BD-21 3153, is a member of the young (about
10 Myr) TW Hya association \citep{Makarov+2001}.  This system would
not be considered suitable for the gravity test according to our
criteria.  Regarding pair number 33, composed of Gaia DR3 3116331104937277952 and 3116324881524878208, the latter star is classified in SIMBAD as an emission-line F9 star. We do not have access to the original LAMOST spectra; however, in the study presenting the low-resolution LAMOST data \citep{Zhang+2023}, the H$\alpha$ equivalent width is reported as negative ($-1.74$ \AA), corresponding to absorption in their convention. This suggests that the star may have been misclassified as an emission-line object. If this is the case, this WB would be fully eligible for the gravity test.  Finally, for the last system (pair~3 in their list), the binary components HD~91234 and HD~91235 appear entirely normal. According to \citet{Francois1986}, HD~91234 is rather poor in
metals, suggesting that the system is likely very old.  The projected
separation for this binary is 19 kAU, and the $D_1-D_2$ distance is 74 kAU
\citep{Chae+2026}. With a 3D separation $r \sim 76$~kAU, disruption by stellar encounters is very likely
during the stellar lifetime \citep{Hamilton+2024}.   We note, however, that the uncertainty in the separation for this binary is very large and that $D_1=D_2$ is within 1 $\sigma$.
In summary, out of
the 3 new WBs found by \citet{Chae+2026} that cannot be fit by bound
Newtonian orbits, one is a young pair, that we would not consider
suitable for the test, and one has a rather large (76 kAU) separation
that makes it a potential disrupted system; one system (pair 33),
composed of Gaia DR3 3116331104937277952 and 3116324881524878208, appears to be quite normal (assuming the second star is misclassified in SIMBAD). The pair has a projected separation of $r \sim 50$ kAU, which is large but not extreme, and a low $r/s$ ratio.

 Several recent works aimed at testing gravity with WBs:
\citet{Cookson+2026} uses a 2D analysis on a large sample, while
\citet{Saad+2026} analyse the \citet{Chae+2026} sample (3D analysis);
both found that Newtonian dynamics well represent the WB
observations.  In more detail, \citet{Cookson+2026} point out that the
2D analysis results depend heavily on the criteria used to
select the sample, and provide a list of criteria that a good sample
should obey. They also provide an extensive summary of the results on
the 2D gravity tests prior to their publication.  \citet{Chae+2026b}
  discuss the results of this work and conclude that their sample of low acceleration binaries is too small to derive strong conclusions.

\citet{Saad+2026} applied a 3D Bayesian analysis to the sample of
\citet{Chae+2026}, modifying the treatment of how the orbital
solutions are obtained. These authors find that the evidence for non
Newtonian physics in Bayesian analysis is sensitive to how the three-dimensional orbital
separation is modeled. 

Ultimately, we argue that, at the current stage of research, binaries for which no bound Newtonian orbits can be found are crucial for tests of gravity and must be studied in detail. Demonstrating definitively whether they are bound  will  remain extremely challenging; however, larger and cleaner samples will allow us to constrain their occurrence rate and assess whether a disruption scenario is plausible or whether a departure from Newtonian dynamics is required. Larger samples will allow to better constrain the $\tilde{\nu}$ test.

In addition to the gravity test, finding orbital solutions to WBs
provides additional information about binary formation and evolution.
An interesting aspect of our (limited) sample is that we do not find
any evidence for super-thermal distribution of eccentricity
\citep{Hwang+2022}, but our orbital solutions are consistent with a
thermal distribution. This is likely due to our limited sample, but we should recall that our initial sample of WBs was larger, 107
  couples, out of which only 60$\%$ survived  our strict selection criteria.
It is worth noting that also \citet{Chae+2026} use a thermal
distribution as a prior, finding that this is consistent with their
results. A hint may come from the fact that our sample has been
cleaned of multiple systems, `disrupted' binaries and members of young associations,
while \citet{Chae+2026} found evidence for non-Newtonian gravity. It
is therefore possible that the excess of eccentricity reported by
\citet{Hwang+2022} is not a result of binary formation, but could
instead be explained by the same causes -- either non suitable binaries or
new physics -- that affect wide binary samples.

\section{Conclusions}
\label{sec_6}
Using a Monte Carlo (MC) approach, we derived the orbital parameters of 64 WBs with precise radial velocity measurements,
including main-sequence solar-type stars previously analysed for gravity tests \citep{Saglia+2025, Pasquini+2026},
as well as a new sample of low-mass M-dwarf twins from a completely independent study \citep{Couperus+2025a, Couperus+2025b}. 
The MC analysis provides the solution space and we use it to verify whether the behavior of the orbital parameters is the expected one,
i.e., the constant distribution of the periastron angle, the orbital inclination proportional to $\sin(i)$, the 
thermal eccentricity ($f(e) = 2e$) and the phase as expected by the thermal eccentricity.

When dividing the sample in two subsamples of 'high' and 'low' acceleration  Eccentricity and periastron follow the predictions;
inclination and phase are distorted by the separation uncertainty,
more strongly for the high-acceleration subsample, which is the opposite of what a gravitational signal would produce. This holds independently
of the threshold value chosen.
  Our analysis is however limited by the error in parallax, so by our poor knowledge of the distance between the stars.    
  For this reason it is extremely powerful to use diagnostics that do not involve the 3D separation between the stars.

  By using $\tilde{\nu}$, that involves very precise observational quantities (total velocity V$_{tot}$, mass and  projected separation s) we do not find evidence of
  excess velocity at low accelerations; and by inserting a boosting factor $\gamma$ in velocity, our limited sample limits it for the low acceleration subsample between
  0.93 and 1.03 at 1 $\sigma$ (please note that this $\gamma$ corresponds to the square root of the gravity boost factor used by \citet{Chae+2026}. 

For the majority of the systems, the MC analysis yields solutions fully consistent with bound Newtonian orbits.
However, five binaries cannot be fitted with bound Newtonian solutions. For four of them we find evidence consistent with their classification as disrupted binaries, following the predictions of works studying the secular evolution of wide binaries. These studies  show that wide binaries 
cannot be treated as isolated systems for their lifetime, and  that the interaction with Galactic tides and stellar encounters produces
a non negligible number of WBs in semi-periodic orbits \citep{Jiang+2010, Hamilton+2024}.
The fifth system is composed of a pair of M stars belonging to a very young association. 

We therefore conclude that Newtonian gravity provides so far a coherent
description of the  observations, and we do not see compelling evidence to abandon Newtonian dynamics to explain wide binaries. 

We acknowledge that other studies report different results, and we discuss why a direct comparison with our work
is not straightforward, due to significant differences in both the samples and the analysis. We believe that the key question is whether systems with large velocity differences (exceeding the Newtonian escape velocity) are truly bound or disrupted (yet with the two stars still relatively close).
We show that a careful analysis of these systems is crucial, and we recommend that binaries belonging to young moving groups, clusters,
and associations should be excluded when testing gravity with wide binaries. 

An interesting result of our MC analysis is also that the space of the orbital solutions parameters is complex, and cannot be
represented by the `best  solution' with an associated Gaussian uncertainty.

Finally, we do not confirm the super-thermal distribution of eccentricities reported in {\it Gaia} projected velocities by \citet{Hwang+2022}.
Our sample is small, but it remains the only one for which strict selection criteria were used and all orbital parameters are fully constrained.

\begin{acknowledgements}
  The authors thank an anonymous referee for several suggestions that improved the quality of the paper.
  L.P.  used  the Gemini and Claude  AI for coding python scripts. 
  Based on data from the European Space Agency (ESA) {\it Gaia} mission, processed by the {\it Gaia} Data Processing and Analysis Consortium (DPAC), https://www.cosmos.esa.int/web/gaia/dpac/consortium.
 L.B. and L.M. acknowledge support from INAF through the Large Grants EPOCH and WST, funding for the WEAVE project, the Mini-Grants Checs , the PRIN PNRR Project Cosmic POT (2022X4TM3H, MUR). J.C. acknowledges support from ANID via Proyecto Fondecyt Regular 1231345, and BASAL project FB210003. Research activities of the observational astronomy board at the Federal University of Rio Grande do Norte (NAOS) are supported by continuous grants from the Brazilian funding agency CNPq. J.R.M. acknowledges a CNPq research fellowship. L. P. acknowledges travel support from the CNPq (Grant 408100/2025-7). B.L.C.M. acknowledges CNPq research fellowships (grant no. 305804/2022-7) and Universal (grant no. 408100/2025-7).
\end{acknowledgements}

\bibliographystyle{aa} 
\bibliography{wb4} 

\begin{appendix}
\section{Orbit determination and uncertainty propagation}
\label{sec:orbit-long}

The orbital elements of each pair are obtained by a Monte Carlo procedure,
organised in two distinct steps.  In the first, described in
Sect.~\ref{sec:closed-form}, we exploit the fact that a wide binary observed
at a single epoch provides a complete relative state vector: given the
distances, the total mass and the relative velocity, the six Keplerian
elements are determined \emph{uniquely} and in closed form, so that no fitting,
optimisation or iterative search is required.  In the second, described in
Sect.~\ref{sec:sampling}, the observational uncertainties are propagated by
generating a large ensemble of Monte Carlo realisations of the six uncertain
observables, solving the orbit exactly for each realisation.
Typical runs comprise $N\sim10^{5}$--$10^{6}$ realisations per pair,
which the closed-form solution makes computationally inexpensive.

\subsection{Reference frame and conventions}
\label{sec:frame}

All orbital quantities are expressed in the local
North--East--line-of-sight frame attached to the primary,
\begin{equation}
  \left(\hat{\mathbf{e}}_{\delta},\,
        \hat{\mathbf{e}}_{\alpha},\,
        \hat{\mathbf{e}}_{\rm los}\right),
\end{equation}
where $\hat{\mathbf{e}}_{\delta}$ points North, $\hat{\mathbf{e}}_{\alpha}$
points East, and $\hat{\mathbf{e}}_{\rm los}$ points from the observer towards
the system.  Explicitly, for a target at $(\alpha,\delta)$,
\begin{align}
  \hat{\mathbf{e}}_{\rm los} &=
    (\cos\delta\cos\alpha,\;\cos\delta\sin\alpha,\;\sin\delta), \\
  \hat{\mathbf{e}}_{\alpha}  &= (-\sin\alpha,\;\cos\alpha,\;0), \\
  \hat{\mathbf{e}}_{\delta}  &=
    (-\cos\alpha\sin\delta,\;-\sin\alpha\sin\delta,\;\cos\delta),
  \label{eq:basis}
\end{align}
the components being given in ICRS.  We stress that this frame is
\emph{local}: its axes are constructed from the ICRS directions evaluated at
the position of the primary and therefore vary across the sky, in contrast to
the fixed equatorial axes of the ICRS itself.  The relative velocity of the
pair is formed as the difference of the two full ICRS space velocities and only
subsequently projected onto this frame, so that the  difference
in the line-of-sight direction towards the two components --- the perspective
term, is  not negligible for some of our systems (largest effect is in WB \#5 from \citet{Saglia+2025}, for an  amount of 178 m\,s$^{-1}$ or 29$\%$ of the observed difference in velocity) is correctly
retained. This has been verified: assigning both components the same 3D space velocity (i.e. projecting it onto each star’s line of sight, which for WB \#5
  changes the proper motions by up to 2 mas/yr and the RV by 0.11 km\,s$^{-1}$) returns a relative velocity of zero, whereas assigning identical observed proper motions and RVs returns
  178 m\,s$^{-1}$, the size of the perspective effect. Writing a relative position as
$\mathbf{R}=(x,y,z)$ in this triple, the standard visual-binary relations hold,
\begin{align}
  x &= r\left[\cos\phi\cos\Omega
       -\sin\phi\cos i\,\sin\Omega\right], \\
  y &= r\left[\cos\phi\sin\Omega
       +\sin\phi\cos i\,\cos\Omega\right], \\
  z &= r\,\sin\phi\sin i ,
\end{align}
with $r=\lvert\mathbf{R}\rvert$ and $\phi-\phi_0$ the true anomaly.  With this
convention $i=0^\circ$ is a face-on  orbit whose motion
proceeds towards increasing position angle, i.e.\ from North towards East;
$i=90^\circ$ is edge-on; and $90^\circ<i\le180^\circ$ denotes an orbit that is
retrograde as projected on the sky.  The longitude of the ascending node
$\Omega$ is the position angle, measured North through East, of the node at
which the companion is receding ($\dot z>0$), and $\phi_0$ is measured within
the orbital plane from that node to periastron in the direction of motion.

We do \emph{not} fold retrograde solutions into $i\in[0^\circ,
90^\circ]$.  The mapping $(i,\Omega,\phi_0)\rightarrow(\pi-i,\Omega+\pi,
\phi_0+\pi)$ is not a symmetry of the orbit: it alters the projected position
on the sky.  The only exact degeneracy of an astrometry-only orbit is
$(\Omega,\phi_0)\rightarrow(\Omega+\pi,\phi_0+\pi)$, which leaves $x$ and $y$
invariant and reverses the sign of $z$; because our analysis incorporates
radial velocities, this degeneracy is broken and $i>90^\circ$ carries genuine
physical information.

Finally, we adopt $\mathbf{R}=\mathbf{r}_1-\mathbf{r}_2$ and
$\mathbf{V}=\mathbf{v}_1-\mathbf{v}_2$, so the elements describe the orbit of
star~1 about star~2.  The inclination and the node are unaffected by this
choice, whereas $\phi_0$ differs by $180^\circ$ from the conventional
secondary-about-primary value.

\subsection{Closed-form determination of the elements}
\label{sec:closed-form}

For a fixed choice of the distances $D_1,D_2$, the total mass $M=m_1+m_2$ and
the relative velocity, the relative position and velocity
$(\mathbf{R},\mathbf{V})$ are fully specified, and the Keplerian elements are
then determined \emph{uniquely}.  No fitting is required.  With $\mu=GM$,
$r=\lvert\mathbf{R}\rvert$ and $V^2=\mathbf{V}\cdot\mathbf{V}$, the specific
orbital energy gives the semi-major axis,
\begin{equation}
  \varepsilon=\tfrac{1}{2}V^{2}-\frac{\mu}{r},
  \qquad
  a=-\frac{\mu}{2\varepsilon},
  \label{eq:vis-viva}
\end{equation}
which is defined only for bound configurations, $\varepsilon<0$.  The specific
angular momentum, the node vector and the eccentricity vector,
\begin{equation}
  \mathbf{h}=\mathbf{R}\times\mathbf{V},
  \qquad
  \mathbf{n}=\hat{\mathbf{z}}\times\mathbf{h},
  \qquad
  \mathbf{e}=\frac{1}{\mu}\left[\left(V^{2}-\frac{\mu}{r}\right)\mathbf{R}
    -(\mathbf{R}\cdot\mathbf{V})\,\mathbf{V}\right],
  \label{eq:vectors}
\end{equation}
then yield the eccentricity $e=\lvert\mathbf{e}\rvert$ and the three angles,
\begin{align}
  i      &= \arccos\!\left(h_z/\lvert\mathbf{h}\rvert\right), \\
  \Omega &= \operatorname{atan2}(n_y,\,n_x), \\
  \phi_0 &= \operatorname{atan2}\!\left(
\mathbf{e}\cdot(\hat{\mathbf{h}}\times\hat{\mathbf{n}}),\;
              \mathbf{e}\cdot\hat{\mathbf{n}}\right), \\
  \phi-\phi_0 &= \operatorname{atan2}\!\left(
              \mathbf{R}\cdot(\hat{\mathbf{h}}\times\hat{\mathbf{e}}),\;
              \mathbf{R}\cdot\hat{\mathbf{e}}\right),
\end{align}
all reduced to $[0,2\pi)$.  The eccentric and mean anomalies follow from
\begin{equation}
  E=2\arctan\!\left(\sqrt{\frac{1-e}{1+e}}\,\tan\frac{\phi-\phi_0}{2}\right),
  \qquad
  M_0=E-e\sin E ,
\end{equation}
and the period from Kepler's third law, $P=\sqrt{a^{3}/M}$ with $a$ in au,
$M$ in solar masses and $P$ in yr.  The degenerate limits are handled explicitly:
for $\lvert\mathbf{n}\rvert\rightarrow0$ (exactly face-on) $\Omega$ is
undefined and set to zero, and for $e\rightarrow0$ (circular) $\phi_0$ is
undefined and set to zero, with $\phi-\phi_0$ then measured from the node.  Both cases
have measure zero under continuous sampling.

We also record the argument of latitude $\phi$, i.e. the sum of the argument of
periastron and the true anomaly.  It is worth noting
that the frequently used geometric relation between the projected and true
separations,
\begin{equation}
  \left(\frac{s}{r}\right)^{2}=1-\sin^{2}\phi\,\sin^{2}i ,
  \label{eq:s-over-r}
\end{equation}
determines only $\lvert\cos\phi\rvert$ and \emph{not} the quadrant of $\phi$.
We therefore use Eq.~(\ref{eq:s-over-r}) solely as an internal consistency
check on $\lvert\cos\phi\rvert$, and never to reconstruct $\phi$
itself.

\subsection{Monte Carlo propagation of the measurement uncertainties}
\label{sec:sampling}

Since the elements are determined exactly by
$(D_1,D_2,M,\mathbf{V}_{\rm rel})$, the observational uncertainties are
propagated by sampling rather than by optimisation.  Six quantities carry
uncertainty: the two parallaxes $\varpi_1,\varpi_2$, the total mass $M$, and
the three components of the relative velocity.  We parameterise each by its
standardised residual,
\begin{equation}
  z_k=\frac{q_k-q_k^{\rm cat}}{\sigma_{q_k}},
\end{equation}
and generate the Monte Carlo ensemble by drawing $\mathbf{z}$ \emph{uniformly}
over the box $[-n_\sigma,+n_\sigma]^{6}$ with $n_\sigma=3$.

\begin{equation}
  \chi^{2}_{\rm tot}=\lvert\mathbf{z}\rvert^{2}
    =\underbrace{z_{\varpi_1}^{2}+z_{\varpi_2}^{2}}_{\chi^{2}_{D_1D_2}}
    +\underbrace{z_{M}^{2}}_{\chi^{2}_{M}}
    +\underbrace{z_{V_\alpha}^{2}+z_{V_\delta}^{2}+z_{V_r}^{2}}_{\chi^{2}_{V}},
  \label{eq:chi2tot}
\end{equation}

Only the relative velocity is sampled, rather than the six per-star
observables.  The orbit depends on $\mathbf{V}_{\rm rel}$ alone; the common
space motion of the pair spans a three-dimensional null space which the
elements do not probe, and including it would reduce the effective sample size
by roughly an order of magnitude without adding information.  The sampling is
nested: $N_{\rm pos}$ draws of $(\varpi_1,\varpi_2,M)$, each combined with
$N_{\rm vel}$ draws of the relative velocity, for
$N=N_{\rm pos}N_{\rm vel}$ realisations in total.  Draws yielding a
non-physical configuration ($\varpi\le0$ or $M\le0$) are discarded, as are
those exceeding a threshold $\chi^{2}_{\rm max}$ (default $30$, corresponding
to $w<4\times10^{-7}$).

Each realisation is solved with the closed-form expressions of
Sect.~\ref{sec:closed-form}.  Realisations that are unbound
($\varepsilon\ge0$), radial ($\lvert\mathbf{h}\rvert=0$) or non-elliptical
($e\ge1$) are rejected, and the corresponding fractions are recorded, since a
low acceptance rate indicates that the resulting distributions are shaped
predominantly by the bound-orbit requirement rather than by the measurements.

For each accepted realisation we record the elements
$(a,e,i,\Omega,\phi_0,\phi-\phi_0)$, the argument of latitude $\phi$, the period
through $\log P$, and the Newtonian acceleration
\begin{equation}
  \log a_{\rm N}=\log_{10}\!\left(\frac{GM}{r^{2}}\right),
\end{equation}
together with the sampled configuration
$(D_1,D_2,M,\mathbf{V}_{\rm rel})$ and the corresponding
$\chi^{2}$ terms of Eq.~(\ref{eq:chi2tot}).

\subsection{Implementation and verification}
\label{sec:verification}

The solver is fully vectorised over the sample axis, and the per-realisation
coordinate transformations are evaluated in closed form rather than through
generic coordinate machinery, which allows $\sim\!10^{6}$ realisations per pair
to be processed in seconds.  Correctness is established at three levels.

\emph{(i) Internal consistency.}  Every solution is propagated back through a
forward Keplerian model, and the residual between the reconstructed and the
sampled state vector is monitored.  Over $3\times10^{3}$ random orbits spanning
the full range of eccentricity and inclination, and including the round trip
through ICRS and back, the maximum relative residuals were
$\lvert\Delta\mathbf{R}\rvert/\lvert\mathbf{R}\rvert=6\times10^{-13}$ and
$\lvert\Delta\mathbf{V}\rvert/\lvert\mathbf{V}\rvert=3\times10^{-13}$.  The
true anomaly is computed by two independent routes --- from the closed-form
inversion, and from $\mathbf{R}\cdot\mathbf{V}$ together with the vis-viva and
angular-momentum relations --- and the two agree to $<10^{-9}$~deg.  The
identity of Eq.~(\ref{eq:s-over-r}) is verified to $<10^{-11}$.

\emph{(ii) External validation.}  The elements were validated against direct
numerical integration of the two-body problem.  Starting from a random state
vector, the equation of motion
$\ddot{\mathbf{R}}=-\mu\mathbf{R}/\lvert\mathbf{R}\rvert^{3}$ was integrated
over a full period in ICRS with an eighth-order Runge--Kutta scheme
(relative tolerance $10^{-12}$), and the elements were read directly from the
resulting trajectory using their physical definitions, without recourse to any
Keplerian formula: the eccentricity from
$(r_{\rm max}-r_{\rm min})/(r_{\rm max}+r_{\rm min})$; the inclination from the
minimum of $s(t)/r(t)$ over the orbit together with the sense of the change of
position angle; the node from the epoch at which $z$ crosses zero with
$\dot z>0$; and $\phi_0$ and $\phi-\phi_0$ from the angles swept within the orbital
plane in the direction of motion.  Over twelve independent trials spanning
$0.15\le e\le0.92$ and both prograde and retrograde orientations, the maximum
discrepancies were $5.8\times10^{-11}$ in $e$ and $8.4\times10^{-7}$~deg in the
angles, consistent with the tolerance of the integrator and of the root finding
rather than with any systematic error.

\emph{(iii) Kinematic cross-check.}  The closed-form ICRS space velocity used
in place of generic coordinate transformations is verified at run time against
the standard implementation, agreeing to better than
$10^{-12}~\mathrm{km\,s^{-1}}$.

\begin{table}
\centering
\caption{Principal parameters of the sampling.}
\label{tab:params}
\begin{tabular}{llc}
\hline\hline
Symbol & Meaning & Default \\
\hline
$N_{\rm pos}$      & draws of $(\varpi_1,\varpi_2,M)$          & $2000$ \\
$N_{\rm vel}$      & velocity draws per position draw          & $250$  \\
$n_\sigma$         & box half-width, mass and velocity         & $3$    \\
$n_\sigma^{\varpi}$& box half-width, parallaxes                & $3$    \\
$\chi^{2}_{\rm max}$ & retention threshold on Eq.~(\ref{eq:chi2tot}) & $30$ \\
\hline
\end{tabular}
\end{table}

\section{Tables}

\onecolumn
\setlength{\tabcolsep}{2pt}
\small
\begin{longtable}{@{} l l l l c c c c c c c l @{}}
\caption{Improved stellar parameters and radial velocities for the highly deviating stars of \citet{Pasquini+2026}. The numbering system is the same as in the \citet{Pasquini+2026} paper.}
\label{tab:stellar_par}\\
\hline\hline
No & Star & Age & T$_{\rm eff}$ & $\log g$ & [Fe/H] & Mass & Radius & RV$_{\rm gr}$ & RV$_{\rm con}$ & Mask/N & RV$_E$ \\
   &      & (Gyr) & (K)           &          &        & (M$_\odot$) & (R$_\odot$) & (km~s$^{-1}$) & (km~s$^{-1}$) &        & (km~s$^{-1}$) \\
\hline
\endfirsthead
\caption{continued.}\\
\hline\hline
No & Star & Age & T$_{\rm eff}$ & $\log g$ & [Fe/H] & Mass & Radius & RV$_{\rm gr}$ & RV$_{\rm con}$ & Mask/N & RV$_E$ \\
\hline
\endhead
\hline
\endfoot
\hline
\multicolumn{12}{l}{\rule{0pt}{3ex}\small \textit{Notes.} RV$_E$ is the radial velocity as measured by ESPRESSO.} \\
\endlastfoot
1  &  TYC 7527-458-1  &   4.7   &   6455 $\pm$ 59   &   4.05 $\pm$ 0.05  &  -0.40 $\pm$ 0.02  &  1.11  &  1.65    &     0.43 &  -0.88 &  F9/6 &  -9.116  $\pm$ 0.003\\        
1  &  TYC 7527-959-1  &   2.5  &   6835 $\pm$ 65   &  4.21 $\pm$ 0.05  &  -0.37 $\pm$ 0.02    &  1.21  &  1.46    &     0.53 &  -0.91 &  F9/4 &  -8.980  $\pm$ 0.021\\      
2  &  TYC 8040-427-1  &  2.9  &   6436 $\pm$ 57  &  4.31 $\pm$ 0.05    &  -0.21 $\pm$ 0.02   &  1.13  &  1.23    &     0.58 &  -0.83 &  F9/7 &  32.957  $\pm$ 0.005\\      
2  &  TYC 8040-648-1  &  3.1  &   6426 $\pm$ 57   &  4.31 $\pm$ 0.05   &  -0.20 $\pm$ 0.02   &  1.12  &  1.25    &     0.57 &  -0.82 &  F9/6 &  32.826  $\pm$ 0.007\\      
5  &  TYC 9492-248-1  &  7.9   &   6309 $\pm$ 58  &  4.23 $\pm$ 0.05  &  -0.61 $\pm$ 0.02    &  0.92   &  1.25   &     0.47 &  -0.69 &  G2/3 &  18.534  $\pm$ 0.001\\      
5  &  TYC 9492-241-1  &  10.3   &   6140 $\pm$ 56  &  4.21 $\pm$ 0.05  &  -0.60 $\pm$ 0.02   &  0.86   &  1.22   &     0.45 &  -0.68 &  G2/3 &  18.467  $\pm$ 0.002\\      
6  &  HD 294014       &  4.8    &   6459 $\pm$ 58   &  4.17 $\pm$ 0.05  &  -0.39 $\pm$ 0.02  &  1.07  &  1.42    &     0.48 &  -0.88 &  F9/5 &  23.755  $\pm$ 0.002\\      
6  &  HD 294012       &  4.9    &   6446 $\pm$ 58   &  4.17 $\pm$ 0.05  &  -0.39 $\pm$ 0.01  &  1.07  &  1.41    &     0.48 &  -0.85 &  F9/5 &  23.819  $\pm$ 0.004\\      
11 &  TYC 6382-243-1  &  4.3  &   6457 $\pm$ 57   &  4.15 $\pm$ 0.05  &  -0.30 $\pm$ 0.02    &  1.12   &  1.50   &     0.48 &  -0.86 &  F9/5 &  16.659  $\pm$ 0.004\\      
11 &  TYC 6382-241-1  & 3.8  &   6492 $\pm$ 60   &  4.21 $\pm$ 0.05  &  -0.32 $\pm$ 0.02    &  1.12   &  1.39   &     0.51 &  -0.84 &  F9/3 &  16.296  $\pm$ 0.004\\    
\hline 
\end{longtable}

\begin{table*}
\caption{Observed parameters of the M stars sample: Total Mass, and velocity differences between the binary components. $V_{cir}  = \sqrt{GM/s}$ is the circular Newtonian velocity.}
\label{tab:velocities}
\centering
\small
\begin{tabular}{r l c c c c c c c}
\hline\hline
No & Suffix & Mass & $s$ & $V_{\alpha}$ & $V_{\delta}$ & $V_{\rm rad}$ & $V_{\rm tot}$ & $V_{cir}$ \\
 & & ($M_\odot$) & (AU) & (km s$^{-1}$) & (km s$^{-1}$) & (km s$^{-1}$) & (km s$^{-1}$) & (km s$^{-1}$) \\
\hline
1  & RTW0143-0151A+B     & 0.43 & 229  & -0.066 & 1.143  & 0.331  & 1.192 & 1.291 \\
2  & 2MA0201+0117A+B     & 1.07 & 515  & -0.251 & -0.581 & -0.751 & 0.982 & 1.358 \\
3  & RTW0231-5432A+B     & 0.53 & 235  & -0.044 & -0.162 & -0.540 & 0.566 & 1.414 \\
4  & LP719-37A+B          & 0.39 & 163  & 0.588  & -0.821 & 0.371  & 1.076 & 1.457 \\
5  & LP368-99A+B          & 0.66 & 150  & -0.554 & -0.527 & -0.349 & 0.840 & 1.976 \\
6  & L533-2A+B            & 0.79 & 122  & -0.754 & -0.569 & -0.130 & 0.954 & 2.397 \\
7  & LP551-62A+B          & 0.76 & 1030 & -0.155 & 0.277  & 0.072  & 0.326 & 0.809 \\
8  & RTW1133-3447A+B      & 0.80 & 5081 & 0.149  & -0.015 & 0.066  & 0.163 & 0.374 \\
9 & RTW1336-3212A+B      & 0.37 & 113  & -0.757 & -0.638 & 0.611  & 1.163 & 1.704 \\
10 & L197-165A+B          & 0.67 & 2360 & -0.260 & -0.120 & 0.405  & 0.496 & 0.502 \\
11 & GJ1183A+B            & 0.43 & 229  & 0.161  & -0.900 & 0.029  & 0.915 & 1.291 \\
12 & RTW1433-6109A+B      & 0.86 & 312  & -0.071 & -0.187 & 0.299  & 0.360 & 1.564 \\
13 & L1197-68A+B          & 0.77 & 388  & -0.412 & -0.595 & 0.301  & 0.784 & 1.327 \\
14 & L1198-23A+B          & 0.79 & 1522 & -0.638 & 0.019  & 0.449  & 0.780 & 0.679 \\
15 & RTW1512-3941A+B      & 1.07 & 143  & 1.347  & 1.111  & 1.370  & 2.219 & 2.576 \\
16 & NLTT44989A+B         & 0.50 & 87   & 0.683  & -0.012 & -1.640 & 1.776 & 2.258 \\
17 & RTW1812-4656A+B      & 0.85 & 4868 & -0.042 & -0.147 & -0.010 & 0.153 & 0.394 \\
18 & GJ745A+B             & 0.75 & 1007 & 0.093  & -0.678 & 0.362  & 0.774 & 0.813 \\
19 & RTW2011-3824A+B      & 0.63 & 131  & 1.434  & 0.750  & -0.980 & 1.891 & 2.066 \\
20 & RTW2211+0058A+B      & 0.55 & 168  & -0.891 & -0.312 & -0.700 & 1.175 & 1.704 \\
21 & RTW2235+0032A+B      & 0.55 & 181  & 0.758  & -0.955 & 0.940  & 1.540 & 1.642 \\
22 & RTW2241-1625A+B      & 1.12 & 187  & 0.567  & -0.893 & -0.560 & 1.197 & 2.305 \\
23 & L718-71A+B           & 0.95 & 1469 & -0.259 & 0.172  & 0.123  & 0.335 & 0.757 \\
24 & RTW2311-5845A+B      & 0.47 & 198  & 0.162  & 0.376  & -1.260 & 1.324 & 1.451 \\
\hline
\end{tabular}
\end{table*}

\begin{table*}
\caption{Orbital parameters for best-fit solutions for the 24 M stars WBs (minimum $\chi^2_{tot} $).}
\label{tab:orbital_params}
\centering
\small
\begin{tabular}{r l c c c c c c c c c}
\hline\hline
No & $D_1$ & $D_2$ & $i$ & $e$ & $\phi-\phi_0$ & $r$ & $a$ & log $P$ & log $A_N$ & $\chi^2_{tot}$ \\
 & (pc) & (pc) & (deg) & & (deg) & (AU) & (AU) & (yrs) & (m s$^{-2}$) & \\
\hline
1 & 47.009 & 47.008 & 111.6 & 0.87 & 115.7 & 396.4 & 1037.8 & 4.7 & -7.80 & 3.31 \\
2 & 49.269 & 49.275 & 68.7 & 0.92 & 146.7 & 1297.9 & 2011.0 & 4.9 & -8.42 & 0.83 \\
3 & 23.007 & 23.020 & 85.0 & 1.00 & 147.0 & 2665.9 & 53210.1 & 7.2 & -9.35 & 1.85 \\
4 & 26.277 & 26.276 & 65.7 & 0.13 & 226.5 & 280.2 & 259.9 & 3.8 & -7.53 & 0.42 \\
5 & 32.623 & 32.617 & 91.9 & 0.68 & 276.1 & 1222.2 & 2450.6 & 5.2 & -8.58 & 0.34 \\
6 & 29.486 & 29.480 & 86.6 & 0.64 & 354.2 & 1244.4 & 3458.6 & 5.4 & -8.52 & 1.71 \\
7 & 40.960 & 41.010 & 89.5 & 0.39 & 314.4 & 10257.0 & 15435.1 & 6.3 & -10.37 & 0.35 \\
8 & 47.899 & 47.902 & 119.8 & 0.95 & 175.4 & 5124.1 & 2785.3 & 5.2 & -9.74 & 0.16 \\
9 & 23.942 & 23.940 & 105.0 & 0.76 & 65.3 & 421.0 & 1326.6 & 4.9 & -7.91 & 1.45 \\
10 & 30.244 & 30.211 & 79.8 & 0.93 & 52.4 & 7205.0 & 86563.2 & 7.5 & -10.11 & 0.66 \\
11 & 17.541 & 17.538 & 101.5 & 0.55 & 59.6 & 651.0 & 1195.2 & 4.8 & -8.23 & 0.35 \\
12 & 30.634 & 30.647 & 96.1 & 0.88 & 197.6 & 2853.7 & 2022.9 & 5.0 & -9.21 & 0.39 \\
13 & 34.444 & 34.435 & 94.2 & 0.55 & 18.1 & 1812.2 & 3931.2 & 5.4 & -8.86 & 0.27 \\
14 & 33.974 & 33.972 & 113.0 & 0.89 & 137.5 & 1544.3 & 2555.8 & 5.2 & -8.71 & 0.45 \\
15 & 23.606 & 23.608 & 111.6 & 0.94 & 282.9 & 387.6 & 3936.5 & 5.4 & -7.37 & 1.28 \\
16 & 18.297 & 18.297 & 81.2 & 0.82 & 155.1 & 102.7 & 80.5 & 3.0 & -6.55 & 2.91 \\
17 & 45.101 & 45.050 & 68.6 & 0.62 & 178.1 & 11517.6 & 7104.7 & 5.8 & -10.42 & 0.38 \\
18 & 8.830 & 8.833 & 146.3 & 0.59 & 237.5 & 1182.9 & 1244.5 & 4.7 & -8.50 & 0.35 \\
19 & 31.372 & 31.373 & 121.5 & 0.74 & 81.6 & 251.3 & 619.2 & 4.3 & -7.24 & 9.11 \\
20 & 33.664 & 33.662 & 73.7 & 0.76 & 246.0 & 521.7 & 858.2 & 4.5 & -7.92 & 0.49 \\
21 & 41.109 & 41.108 & 136.0 & 0.86 & 141.2 & 251.8 & 322.7 & 3.9 & -7.29 & 1.10 \\
22 & 41.982 & 41.988 & 98.0 & 0.91 & 58.2 & 1314.0 & 11525.8 & 6.1 & -8.41 & 3.65 \\
23 & 19.699 & 19.694 & 117.4 & 0.93 & 173.7 & 1824.7 & 1018.5 & 4.5 & -8.77 & 0.24 \\
24 & 43.843 & 43.837 & 80.4 & 0.98 & 238.0 & 1180.5 & 14747.6 & 6.4 & -8.71 & 7.97 \\
\hline
\end{tabular}
\end{table*}

\begin{table}
\caption{Weighted median $\log A_N$ (m\,s$^{-2}$) used to classify the binaries. Values below the threshold of $-9$ (low acceleration) are marked with an asterisk. Numbering as in \citet{Saglia+2025} (HARPS), \citet{Pasquini+2026} (Chanam\'e) and Table~\ref{tab:velocities} (M stars).}
\label{tab:logAN}
\centering
\begin{tabular}{rr|rr|rr}
\hline\hline
\multicolumn{2}{c|}{HARPS} & \multicolumn{2}{c|}{Chanam\'e} & \multicolumn{2}{c}{M stars} \\
\# & $\log A_N$ & \# & $\log A_N$ & \# & $\log A_N$ \\
\hline
1 & $-9.50$$^*$ & 1 & $-11.92$$^*$ & 1 & $-7.49$ \\
2 & $-8.11$ & 2 & $-12.31$$^*$ & 2 & $-8.28$ \\
3 & $-9.29$$^*$ & 3 & $-12.03$$^*$ & 3 & $-9.18$$^*$ \\
4 & $-8.31$ & 4 & $-11.27$$^*$ & 4 & $-7.53$ \\
5 & $-9.18$$^*$ & 5 & $-11.47$$^*$ & 5 & $-8.25$ \\
6 & $-10.40$$^*$ & 6 & $-12.00$$^*$ & 6 & $-8.22$ \\
7 & $-7.58$ & 8 & $-11.05$$^*$ & 7 & $-9.92$$^*$ \\
8 & $-8.94$ & 9 & $-10.54$$^*$ & 8 & $-10.03$$^*$ \\
9 & $-9.93$$^*$ & 10 & $-10.36$$^*$ & 9 & $-7.58$ \\
10 & $-8.43$ &  &  & 10 & $-9.70$$^*$ \\
11 & $-7.85$ &  &  & 11 & $-8.08$ \\
12 & $-10.98$$^*$ &  &  & 12 & $-9.44$$^*$ \\
13 & $-8.86$ &  &  & 13 & $-8.49$ \\
14 & $-9.05$$^*$ &  &  & 14 & $-8.78$ \\
15 & $-8.77$ &  &  & 15 & $-7.04$ \\
16 & $-6.93$ &  &  & 16 & $-6.81$ \\
17 & $-7.86$ &  &  & 17 & $-10.63$$^*$ \\
18 & $-7.89$ &  &  & 18 & $-8.49$ \\
19 & $-10.24$$^*$ &  &  & 19 & $-7.24$ \\
20 & $-6.28$ &  &  & 20 & $-7.50$ \\
21 & $-8.79$ &  &  & 21 & $-7.29$ \\
22 & $-8.04$ &  &  & 22 & $-7.97$ \\
23 & $-8.72$ &  &  & 23 & $-8.93$ \\
25 & $-9.27$$^*$ &  &  & 24 & $-7.98$ \\
26 & $-9.06$$^*$ &  &  &  &  \\
27 & $-7.55$ &  &  &  &  \\
28 & $-9.93$$^*$ &  &  &  &  \\
29 & $-7.47$ &  &  &  &  \\
30 & $-7.36$ &  &  &  &  \\
31 & $-10.47$$^*$ &  &  &  &  \\
32 & $-7.59$ &  &  &  &  \\
\hline
\end{tabular}
\end{table}

\end{appendix}
\end{document}